\documentclass{aa}
\input{epsf}
\usepackage{graphicx}
\begin{document}

\title{Impacts of a power--law non--thermal electron tail on the ionization 
and recombination rates}

\author{D. Porquet, M. Arnaud, A. Decourchelle}

\offprints{Delphine Porquet,\\
 \email{dporquet@cea.fr}}
\institute{CEA/DSM/DAPNIA, Service d'Astrophysique, C.E. Saclay,
F-91191 Gif-sur-Yvette Cedex, France}
\date{Received March, 6 2001; accepted April 2001}

\def \etal      {et al.\ }

\def \me {\rm m_{e}}

\def \kT  {{\rm k}T}
\def \Eb {E_{\rm b}}
\def \xb {x_{\rm b}}
\def \a {\alpha}
\def \e {\rm e}
\def \K {\rm K}

\def \DI {\rm_{\rm DI}}
\def \I {\rm_{\rm I}}
\def \EA {\rm_{\rm EA}}
\def \DR {\rm_{\rm DR}}
\def \RR {\rm_{\rm RR}}

\def \sRR {\sigma_{\RR}(E)}
\def \sDR {\sigma_{\DR}(E)}

\def \CIM {C_{\I}^{\rm M}(T)}
\def \CDIM {C_{\DI}^{\rm M}(T)}
\def \CEAM {C_{\EA}^{\rm M}(T)}
\def \rrM {\alpha_{\RR}^{\rm M}(T)}
\def \drM {\alpha_{\DR}^{\rm M}(T)}

\def \CIH {C_{\I}^{\rm H}(T,\xb,\a)}
\def \CDIH {C_{\DI}^{\rm H}(T,\xb,\a)}
\def \CEAH {C_{\EA}^{\rm H}(T,\xb,\a)}
\def \rrH {\alpha_{\RR}^{\rm H}(T,\xb,\a)}
\def \drH {\alpha_{\DR}^{\rm H}(T,\xb,\a)}

\def \bi {\beta_{\I}(T,\xb,\a)}
\def \bRR {\beta_{\RR}(\xb,\a)}
\def \bDR {\beta_{\DR}(T, \xb,\a)}

\def \EDR {E_{\DR}}
\def \EI {E_{\I}}
\def \xj {x_{\rm j}}
\def \xi {x_{\rm i}}
\def \ubj {u_{\rm b_{\rm j}}}
\def \xEA {x_{\EA}}
\def \IEA {I_{\EA}}
\def \ucEA {u_{\EA}}

\def \Aj {A_{\rm j}}
\def \Ij {I_{\rm j}}
\def \Bj {B_{\rm j}}
\def \Cj {C_{\rm j}}
\def \Dj {D_{\rm j}}
\def \Uj {U_{\rm j}}
\def \uj {u_{\rm j}}


\abstract{
We have investigated the effects of a non--thermal electron 
population on the ionization and recombination rates. The considered 
electron distribution is defined as a Maxwellian function below a 
break energy $E_{\mathrm b}$ and as a power--law function of index $\alpha$ 
above this energy.
We have calculated the collisional (direct and excitation autoionization)
ionization coefficient rates as well as the (radiative and dielectronic) 
recombination rates. Practical methods are given to calculate these rates 
in order to be easily included in a computer code. The ionization rates 
are very sensitive to the non--thermal electron population and can be 
increased by several orders of magnitude depending on the temperature and 
parameters of the power--law function ($E_{\mathrm b}$ and $\alpha$). 
The non--thermal electrons have a much weaker effect on the (radiative and 
dielectronic) recombination rates. We have determined the mean electric 
charge of elements C, N, O, Ne, Mg, Si, S, Ar, Ca, Fe and Ni for 
different values of the break energy and power--law index. The ionization 
balance is affected significantly, whereas the effect is smaller in ionizing 
plasmas.
}
\titlerunning{Impacts of a power--law electron tail}
\authorrunning{Porquet, Arnaud, Decourchelle}
\maketitle


\section{Introduction}

The ionization and recombination rates for astrophysical plasmas have
usually been calculated for a Maxwellian electron distribution (e.g.,
Arnaud \& Rothenflug \cite{Arnaud85}, Arnaud \& Raymond
\cite{Arnaud92}, Mazzotta \etal \cite{Mazzotta1998}).  However, in
many low-density astrophysical plasmas, electron distributions may
differ from the Maxwellian distribution.  The degree of ionization of
a plasma depends on the shape of the electron distribution, as well as
on the electronic temperature.  This has been studied for the solar
corona (e.g. Roussel-Dupr{\'e} \cite{Roussel-Dupre80}, Owocki \& Scudder \cite{Owocki83}, Dzifc{\'a}kov{\'a} \cite{Dzifcakova92}, Dzifc{\'a}kov{\'a} \cite{Dzifcakova98}) and for evaporating
interstellar clouds (Ballet et al. \cite{Ballet1989}), where a 
non--thermal electron distribution occurs in places where there are high
gradients of density or temperature.

A non--thermal electron population is expected in various astrophysical
plasmas.  Strong shocks can convert a large fraction of their energy
into the acceleration of relativistic particles by the diffusive shock
acceleration process (e.g., Drury \cite{Drury1983}, Blandford \&
Eichler \cite{Blandford1987}, Jones \& Ellison \cite{Jones1991}, Kang
\& Jones \cite{Kang1991}).  Direct evidence for the presence of
accelerated electrons up to relativistic energies ($\simeq~1$GeV)
comes from the observations of radio synchrotron emission in supernova 
remnants and in clusters of galaxies. More recently, non--thermal X-ray 
emission has been reported in several shell-like supernova remnants and
interpreted as synchrotron radiation from cosmic-ray electrons up to
$\simeq~100$ TeV (Koyama et al.  \cite{Koyama1995}, Allen et al. 
\cite{Allen1997}, Koyama et al. 
\cite{Koyama1997}, Slane et al.  \cite{Slane1999}, Slane et al. 
\cite{Slane2001}).

A number of recent works have focused on the non--thermal emission
from supernova remnants (e.g., Laming 2001, Ellison et al. 
\cite{Ellison2000}, Berezhko \& V{\"o}lk \cite{Berezhko2000}, Bykov et 
al.  \cite{Bykov2000b}, Baring et al.  \cite{Baring1999}, Gaisser et al. 
\cite{Gaisser1998}, Reynolds \cite{Reynolds1996,Reynolds1998}, Sturner 
et al. \cite{Sturner1997}) and clusters of galaxies (e.g., Sarazin 
\cite{Sarazin1999}, Bykov et al. \cite{Bykov2000a}, Sarazin \& Kempner 
\cite{Sarazin2000}). The impact of efficient acceleration on the 
hydrodynamics and thermal X-ray emission has been investigated (Decourchelle, Ellison, \& Ballet \cite{Decourchelle2000}, Hughes, Rakowski, \& Decourchelle \cite{Hughes2000}).

When the acceleration is efficient, the non--thermal population is 
expected to modify directly the ionization rates in the plasma as well
as the line excitation (e.g. Dzifc{\'a}kov{\'a} \cite{Dzifcakova2000}, Seely et al. \cite{Seely87}). A hybrid electron distribution (Maxwellian plus power--law tail) is expected from diffusive shock acceleration (e.g.,
Berezhko \& Ellison \cite{Berezhko1999}, Bykov \& Uvarov
\cite{Bykov1999}).  The low energy end of the power--law electron
distribution (which connects to the Maxwellian thermal population) is
likely to enhance the ionization rates and to significantly modify the
degree of ionization of the plasma, which is used as a
diagnostic of the plasma electron temperature.

In this paper, we shall examine the influence of a power--law non--thermal
electron distribution (connecting to the falling Maxwellian thermal
population) on the ionization and recombination rates for C, N, O, Ne, Mg, Si, S, Ar, Ca, Fe and Ni. 
For different characteristic values of the power--law electron
distribution, the mean electric charge of these elements has been
determined as a function of the temperature at ionization equilibrium 
and for different values of the ionization timescale.  
We intend, in this paper, to give a comprehensive study of the 
dependence of these quantities on the parameters of the non--thermal 
population, illustrated by simple examples.  We do not provide tables, 
which would be too numerous as the ionization equilibrium depends in 
our model on four parameters (element, temperature of the thermal 
component, index and low energy break of the non--thermal population).  
In the appendix or directly in the text, we give the formula needed 
for the calculations of the rates which could be easily inserted in 
computer codes.

In Section~\ref{sec:Electrondistributionshapes}, we define the Hybrid 
electron distribution used in this work. The calculation of the new 
ionization collisional rates and (radiative and dielectronic) 
recombination rates is discussed in Section~\ref{sec:rates}. In 
Section~\ref{sec:Ionizationequilibria}, we present the derived mean 
electric charge of the elements in ionization equilibrium as well as 
in ionizing plasmas.  
\section{The electron distribution
shapes}\label{sec:Electrondistributionshapes}

\begin{figure}[t]
\epsfxsize=8.cm \epsfverbosetrue \epsfbox{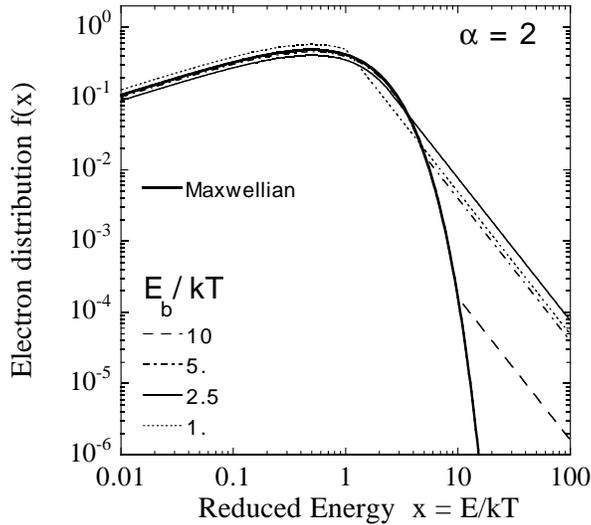} 
\caption{The
Hybrid (Maxwell/Power--law) electron distribution for different values of the
break parameter $\xb$, compared to a Maxwellian distribution
($\xb=\infty$). The slope of the power--law has been fixed to $\a=2$.
Reduced energy coordinates, $x=E/\kT$, are used.}
\label{fig:fdist}
\end{figure}

\begin{figure}[t]
\epsfxsize=8.cm \epsfverbosetrue \epsfbox{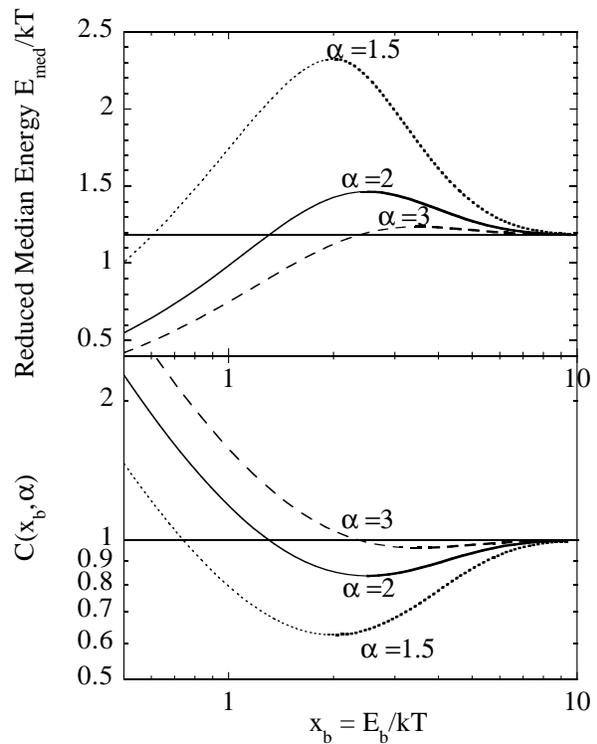}
\caption{Variation of the median energy (top panel) and of the
normalisation constant (bottom panel) with the break parameter $\xb$.  
The lines are labeled by the value of the slope parameter $\a$. The 
bold part of the curves corresponds to $\xb \geq \a+1/2$. The black 
horizontal lines correspond to the values obtained for a Maxwellian 
distribution.}
\label{fig:fCEmed}
\end{figure}

The Maxwellian distribution, generally considered for the electron
distribution in astrophysical plasmas, $N_{\rm e}(E)$, is defined as:
\begin{eqnarray}
    \label{eq:fMaxw}
dN_{\rm e}(E) &= &n_{\rm e}\ f^{\rm M}_{\rm E}(E)\ dE \\
f^{\rm M}_{E}(E)&=&\frac{2}{\sqrt{\pi}}\ ({\rm k}T)^{-3/2}\ E^{1/2}\
e^{-\frac{E}{{\rm k}T}}
\end{eqnarray}
where $E$ is the energy of the electron, $T$ is the electronic 
temperature and $n_{e}$ the total electronic density. In this 
expression the Maxwellian function $f^{\rm M}_{\rm E}(E)$ is 
normalised so that $\int^{\infty}_{0}~f^{\rm M}_{\rm E}(E)~dE=1$.

It is convenient to express this distribution in term of the reduced energy
$x=E/\kT$:
\begin{eqnarray}\label{eq:fM}
dN_{\rm e}(x) &=& n_{\rm e}\ f^{\rm M}(x)\ dx \\
f^{\rm M}(x)&=&\frac{2}{\sqrt{\pi}}\ \ x^{\frac{1}{2}}\ e^{-x}
\label{eq:distM}
\end{eqnarray}
The corresponding scaled (non--dimensional) distribution $f^{\rm
M}(x)$ is an universal function, of fixed shape.  \\

Non--Maxwellian electron distributions expected in the vicinity of
shock waves, as in young supernova remnants, seem to be reasonably 
described by a Maxwellian distribution at low energy up to a break 
energy $\Eb$, and by a power--law distribution  at higher energy
(e.g., Berezhko \& Ellison \cite{Berezhko1999}, Bykov \& Uvarov 
\cite{Bykov1999}).  We call hereafter this ``Maxwellian/Power--law'' 
type of electron distribution the {\bf Hybrid electron distribution} 
($f^{\rm H}$).  It is defined, in reduced energy coordinates, as:

\begin{equation}
    dN_{\rm e}(x) = n_{\rm e}f^{\rm H}(x) dx
\end{equation}
\begin{eqnarray}
    \label{eq:distH}
f^{\rm H}(x)&=&C(x_{\rm b},\a)\ \frac{2}{\sqrt{\pi}}\ x^{1/2}\
e^{-x} ~~~~~~~~~~~~~~~ x \leq x_{\rm b}  \\
f^{\rm H}(x)&=&C(x_{\rm b},\a)\ \frac{2}{\sqrt{\pi}}\
x_{\rm b}^{1/2}\ e^{-x_{\rm b}}\  \left(\frac{x}{x_{\rm b}}\right)^{-\a}~~x \geq  \xb, \nonumber
\end{eqnarray}
where $\xb =\Eb/\kT$ is the reduced break energy, and $\a$ is the energy index of the power--law ($\a>1$). Note that for $\a \leq$ 2~ the energy diverges (in practice a cutoff at very high energy occurs). Since for the calculations of the ionization and recombination rates the very high energies ($\geq$ 20kT) have negligible effect, for simplicity, we use here a power-law defined from $\xb$ to infinity.\\
 The normalisation factor of the  power--law distribution is defined so that the electron distribution is  continous at $\xb$. The factor $C(\xb,\a)$ is a normalisation constant, so that $\int^{\infty}_{0}~f^{\rm H}(x)~dx=1$:
\begin{equation}
 C(x_{\rm b},\a) = \frac{\sqrt{\pi}}{2}\
\frac{1}{\gamma(\frac{3}{2},x_{\rm b}) +
(\a-1)^{-1} \ x_{\rm b}^{3/2}\ e^{-x_{\rm b}}}
 \end{equation}
where $\gamma(a,x)$ is the gamma function defined as
$\gamma(a,x)=\int^{x}_{0}t^{a-1}\ \e^{-t}\ dt$. For $x${\bf $\leq$}
$\xb$, the Hybrid distribution only differs from a Maxwellian
distribution by this multiplicative factor.  The scaled distribution
$f^{\rm H}(x)$ only depends on the two non--dimensional parameters,
$\xb$ and $\a$.  The dependence on $\kT$ of the
corresponding physical electron distribution is $f^{\rm H}_{\rm
E}(E)=(\kT)^{-1} f^{\rm H}(E/\kT)$.\\

The Hybrid distributions $f^{\rm H}(x)$, obtained for several values
of the energy break $\xb$, are compared to the Maxwellian
distribution in Fig.~\ref{fig:fdist}.  The slope has been fixed to
$\a=2$, a typical value found in the models referenced above.  The
variation of the reduced median energy of the distribution with
$\xb$, for $\a=1.5,2.,3.$, is plotted in Fig.~\ref{fig:fCEmed}, as 
well as the variation of the normalisation factor $C(\xb,\a)$.

As apparent in the figures, there is a critical value of $\xb$, for
each $\a$ value, corresponding to a qualitative change in the
behavior of the Hybrid distribution.  This can be understood by
looking at the distribution at the break energy $\xb$.  Whereas the
distribution is continuous, its slope changes.  The logarithmic slope
is $1/2-\xb$ on the Maxwellian side and $-\a$ on the power--law side. 
There is no break in the shape of the Hybrid distribution (full line
in Fig.~\ref{fig:fdist}), only for the critical value of $\xb=\a+1/2$. 
For $\xb>\a+1/2$, the power--law always decreases less rapidly with
energy than a Maxwellian distribution and does correspond to an {\it
enhanced} high energy tail.  The contribution of this tail increases
with decreasing $\xb$ (and $\a$).  Thus the median energy increases
and the normalisation parameter, which scales the Maxwellian part,
decreases (see Fig.~\ref{fig:fCEmed}).  On the other hand, when
$\xb<\a+1/2$, there is an intermediate region above the
energy break where the power--law decreases more steeply than a
Maxwellian (see dotted line in Fig.~\ref{fig:fdist}).  This results in
a deficit of electrons at these energies as compared to a Maxwellian
distribution, more and more pronounced as $\xb$ is small.  The median
energy thus starts to decrease with decreasing $\xb$ and can be even
lower than the median energy of a Maxwellian distribution
(Fig.~\ref{fig:fCEmed}).  In this paper we only consider the
regime where $\xb\geq\a+1/2$.  It corresponds to clear cases where the
high energy part of the distribution is indeed increased, as expected
when electron are accelerated in shocks.  Furthermore, the
distribution used here is only an approximation, valid when the hard
tail can be considered as a perturbation of the original Maxwellian
distribution.  The simulations of Bykov and Uvarov (\cite{Bykov1999},
see their Figure 2) clearly show that the low energy part of the 
distribution is less and less well approximated by a Maxwellian 
distribution, as the `enhanced' high energy tail extends to lower and 
lower energy (lower `break').  Although we cannot rigorously define 
a corresponding quantitative lower limit on $\xb$, the cases presented 
by Bykov and Uvarov (\cite{Bykov1999}, see their Figure 2) suggest a 
limit similar to the one considered here, i.e. a few times the 
Maxwellian peak energy.\\

The distribution considered here differs from the so-called
``kappa-distribution'' or the ``power distribution'', relevant for
other physical conditions (see e.g. Dzifc{\'a}kov{\'a} 2000 and references therein).  These two distributions have been used to
model deviations from a Maxwellian distribution caused by strong
plasma inhomogeneities, as in the solar corona, and their impact on
the ionization balance has been extensively studied (e.g. Roussel-Dupr{\'e} \cite{Roussel-Dupre80}, Owocki \& Scudder \cite{Owocki83}, Dzifc{\'a}kov{\'a} \cite{Dzifcakova92}, Dzifc{\'a}kov{\'a} \cite{Dzifcakova98}).  Although the effect of the Hybrid distribution
is expected to be qualitatively similar, it has never been
quantitatively studied.  In the next section we discuss how the
ionization and recombination rates are modified, as compared to a pure
Maxwellian distribution, depending on the parameters $\xb$ and $\a$.

\begin{figure}[t]
\epsfxsize=8.cm \epsfverbosetrue \epsfbox{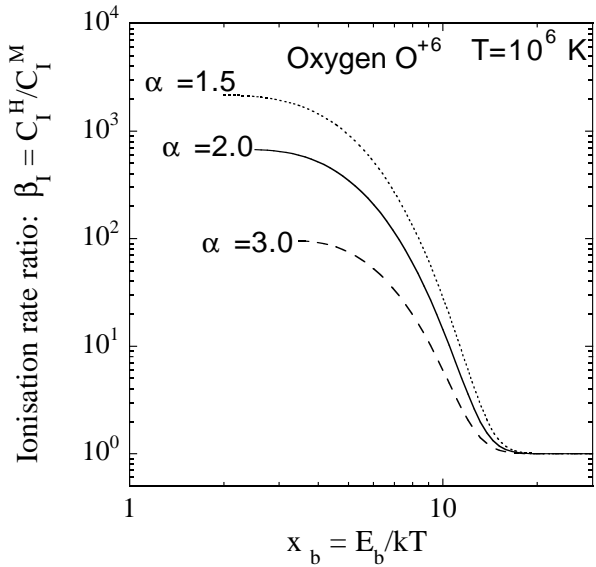}
\caption{Variation of the ratio of the ionization rate in a Hybrid
distribution over that in a Maxwellian with the same temperature,
versus the break parameter $\xb$ of the Hybrid distribution.  The
curves correspond to different values of the slope parameter $\a$ and
are labeled accordingly.  The ion considered is O$^{+6}$, the
temperature is fixed at $10^{6}$ K.}
\label{fig:fIonisO7a}
\end{figure}

\section{Calculations of the collisional ionization  and
recombination rates}\label{sec:rates}

Let us consider a collisional process of cross section $\sigma(E)$,
varying with energy $E$ of the incident electron.  The corresponding
rate coefficient (cm$^{3}$\,s$^{-1}$), either for a Maxwellian distribution or a Hybrid
distribution, $f(x)$, is given by:
\begin{eqnarray}
\mathrm{Rate} &=&\left(\frac{2 {\rm k}T}{\me}\right)^{\frac{1}{2}}\int_{x_{\rm th}}^{\infty}\ x^{\frac{1}{2}}\ \sigma(x {\rm k}T)\ f(x)\ dx
\label{eq:C(kT)}
\end{eqnarray}
with $x_{\rm th}=E_{\rm th}/\kT$. $E_{\rm th}$ corresponds to the threshold energy of the considered process (for $E<E_{\rm th}$, 
$\sigma(E)=0$).  For the recombination processes, no threshold energy 
is involved and $x_{\rm th}=0$.

The rates for the Hybrid distribution depend on $\kT$, $\xb$ and $\a$
and are noted $\CIH$, $\rrH$ and $\drH$ for the ionization, radiative
and dielectronic recombination process respectively.  The
corresponding rates for the Maxwellian distribution which only depends
on $\kT$ are $\CIM$, $\rrM$ and $\drM$.

The ionization data are taken from Arnaud \& Rothenflug
(\cite{Arnaud85}) and Arnaud \& Raymond (\cite{Arnaud92}), as adopted
by Mazzotta \etal (\cite{Mazzotta1998}) for the most abundant elements
considered here.  The recombination data are taken from the updated
calculations of Mazzotta (\cite{Mazzotta1998}).  In the next sections
we outline the general behavior of the rates with the electron
distribution parameters, using mostly oxygen ions (but also iron) as
illustration.


\subsection{The electronic collisional ionization
rates}\label{sec:ionis}
\begin{figure}[t]
\epsfxsize=8.cm \epsfverbosetrue \epsfbox{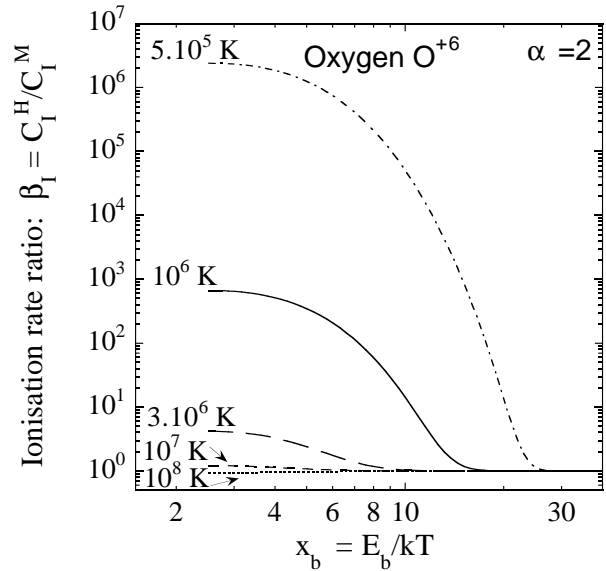}
\caption{Same as Fig.~\ref{fig:fIonisO7a} but the parameter $\a$ is now
fixed to 2. The curves correspond to different values of the
temperature.}
\label{fig:fIonisO7T}
\end{figure}

\begin{figure}[t]
\epsfxsize=8.cm \epsfverbosetrue \epsfbox{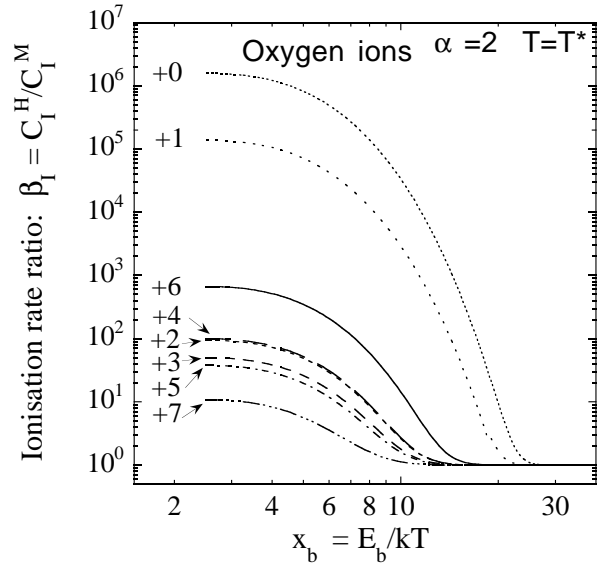}
 \caption{Same as Fig.~\ref{fig:fIonisO7a} for the different ions of oxygen.
Each curve is labeled by the charge of the ion considered.  The slope
parameter is fixed to $\a=2$ and the temperature for each ion is fixed
at the value, $T^{*}$, where the abundance of the ion is maximum for a
Maxwellian electron distribution.}
\label{fig:fIonisOTs}
\end{figure}

The ionization cross sections present a threshold at the first
ionization potential of the ionizing ion, $\EI$.  The cross
sections always present a maximum, at $E_{\rm m}$, and decrease as
$\ln(E)/E$ at very high energies (e.g., Tawara, Kato,\& Ohnishi 1985).  
The ionization rate is very sensitive to the proportion of electrons 
above the threshold and the modification of the ionization rate for the 
Hybrid distribution depends on how the high energy tail affects this 
proportion.

Parametric formulae for the ionization cross sections are available 
from the litterature and it is easy to derive the corresponding rates 
for the Hybrid distribution. This is detailed in Appendix~\ref{app:Ionization}. 

\begin{figure}[t]
\epsfxsize=8.cm \epsfverbosetrue \epsfbox{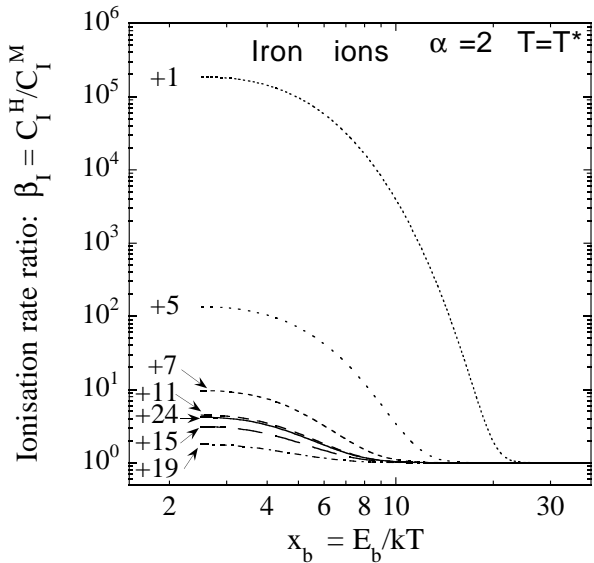}
\caption{Same as Fig.~\ref{fig:fIonisOTs} for the different ions of iron.}
\label{fig:fIonisFeTs}
\end{figure}

\begin{figure*}[t]
\centering
\hspace*{0.3cm}\epsfxsize=12.cm \epsfverbosetrue \epsfbox{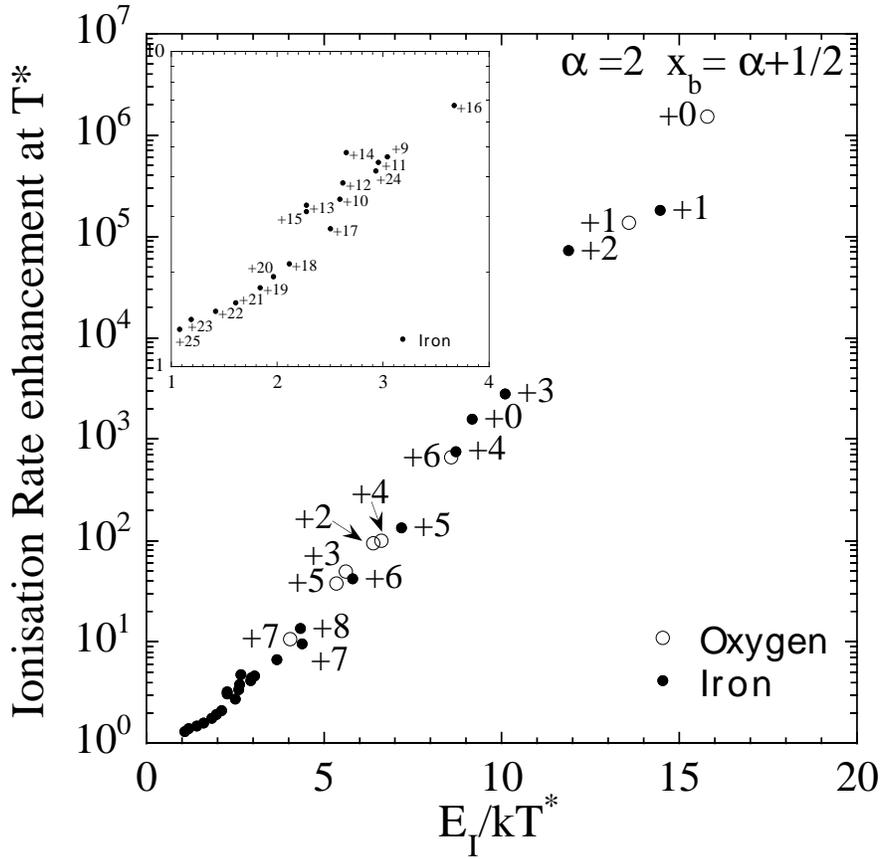}
\caption{Maximum enhancement ratio of the ionization rate, 
$\bi$, at $T^{*}$ versus $\EI/\kT^{*}$.  $T^{*}$ is the
temperature at which the abundance of the ion is maximum for a
Maxwellian electron distribution and $\EI$ is the first
ionization potential of the ion.  Each point corresponds to an ion of
oxygen (open circle) or iron (filled circles).  The points are labeled
by the charge of the corresponding ion.  The slope parameter is fixed
to $\a=2$ and the break is $\xb=\a+\frac{1}{2}$, corresponding to a
maximum enhancement of the ionization rate.}
\label{fig:fIonisratio}
\end{figure*}

To understand the influence of the presence of a high energy power--law
tail in the electron distribution, we computed  the ratio
$\bi=\CIH/\CIM$, of the ionization rate in a Hybrid distribution over
that in a Maxwellian with the same temperature.  This ratio is plotted
in Fig.~\ref{fig:fIonisO7a} to Fig.~\ref{fig:fIonisratio} for
different ions and values of the parameters $\xb$ and $\a$.\\

Let us first consider O$^{+6}$.  Its ionization potential is $E_{\rm
I}=739~{\rm eV}$ and the cross section is maximum at about $3~E_{\rm
I}$.  Its abundance, for a Maxwellian electron distribution, is maximum
at $T^* \simeq 10^{6}$~K under ionization equilibrium (Arnaud \& Rothenflug 
1985).  At this temperature, the threshold energy is well above
the thermal energy ($\EI/\kT\sim 8$) and only the very high
energy tail of the Maxwellian contributes to $\CIM$, i.e. a small
fraction of the electron distribution.  This fraction is dramatically
increased in the Hybrid distribution as soon at the break energy is
not too far off from the threshold, $\xb\sim 15$ for O$^{+6}$
(Fig.~\ref{fig:fIonisO7a}).  The enhancement factor $\bi$ naturally
increases with decreasing break $\xb$ and slope $\a$ parameters
(Fig.~\ref{fig:fIonisO7a}), since the distribution median energy
increases when these parameters are decreased (Fig.~\ref{fig:fCEmed}).  \\

This behavior versus $\xb$ and $\a$ is general at all temperatures as
illustrated in Fig.~\ref{fig:fIonisO7T}, provided that the thermal
energy is not too close to $E_{\rm m}$, i.e. that the majority of the
contribution to the ionization rate is from electrons with energies
corresponding to the increasing part of the ionization cross section. 
If this is no more the case, the ionization rate starts to decrease
with increasing distribution median energy.  Thus, for high enough
values of the temperature (see the curve at $T = 10^{8}$~K in
Fig.~\ref{fig:fIonisO7T}), the factor $\bi$ becomes less than unity
and decreases with decreasing $\xb$.  The correction factor is small
(around $\sim 10\%$) however in that case.\\

More generally the enhancement factor $\bi$ at fixed values of $\xb$
and $\a$, depends on the temperature (Fig.~\ref{fig:fIonisO7T}).  It
decreases with increasing temperature: the peak of the distribution is
shifted to higher energy as the ratio $\kT/\EI$ increases and
the enhancement due to the contribution of the hard energy tail
decreases.\\

The qualitative behavior outlined above does not depend on the ion
considered.  We plotted in Fig.~\ref{fig:fIonisOTs} and in
Fig.~\ref{fig:fIonisFeTs} the enhancement factor for the different
ions of oxygen and a choice of iron ions at $T^{*}$ (the temperature
of maximum ionization fraction of the ion for a Maxwellian electron
distribution under ionization equilibrium). $\EI/\kT^{*}$ is always greater 
than unity and the ionization rates are increased by the Hybrid 
distribution, the enhancement factor $\bi$ increasing with decreasing 
$\xb$.  However this enhancement factor differs from ion to ion, it 
generally increases with increasing $\EI/\kT^{*}$ value (approximatively
with an exponential dependence), as shown in
Fig.~\ref{fig:fIonisratio}.  This is again due to the relative
position of the peak of the distribution with respect to the threshold
energy.  Note that $\EI/\kT^{*}$ is generally smaller for more
ionized ions (but this is not strictly true) so that low charge
species are generally more affected by the Hybrid distribution.  \\

In summary, the Hybrid rates are increased with respect to the
Maxwellian rates except at very high temperature.  The enhancement
factor depends on the temperature, mostly via the factor $\EI/\kT$. 
It increases dramatically with decreasing temperature and is always
important at $T^{*}$, where it can reach several orders of magnitude. 
The ionization balance is thus likely to be affected significantly,
whereas the effect should be smaller in ionizing plasmas but 
important in recombining plasmas.  
For $\xb$ typically lower than 10--20 (with this upper limit higher for lower temperature, see fig.~\ref{fig:fIonisO7T}), the impact of the Hybrid rate increases with decreasing $\xb$ and $\a$.\\

\begin{figure}[t]
\epsfxsize=8.cm \epsfverbosetrue \epsfbox{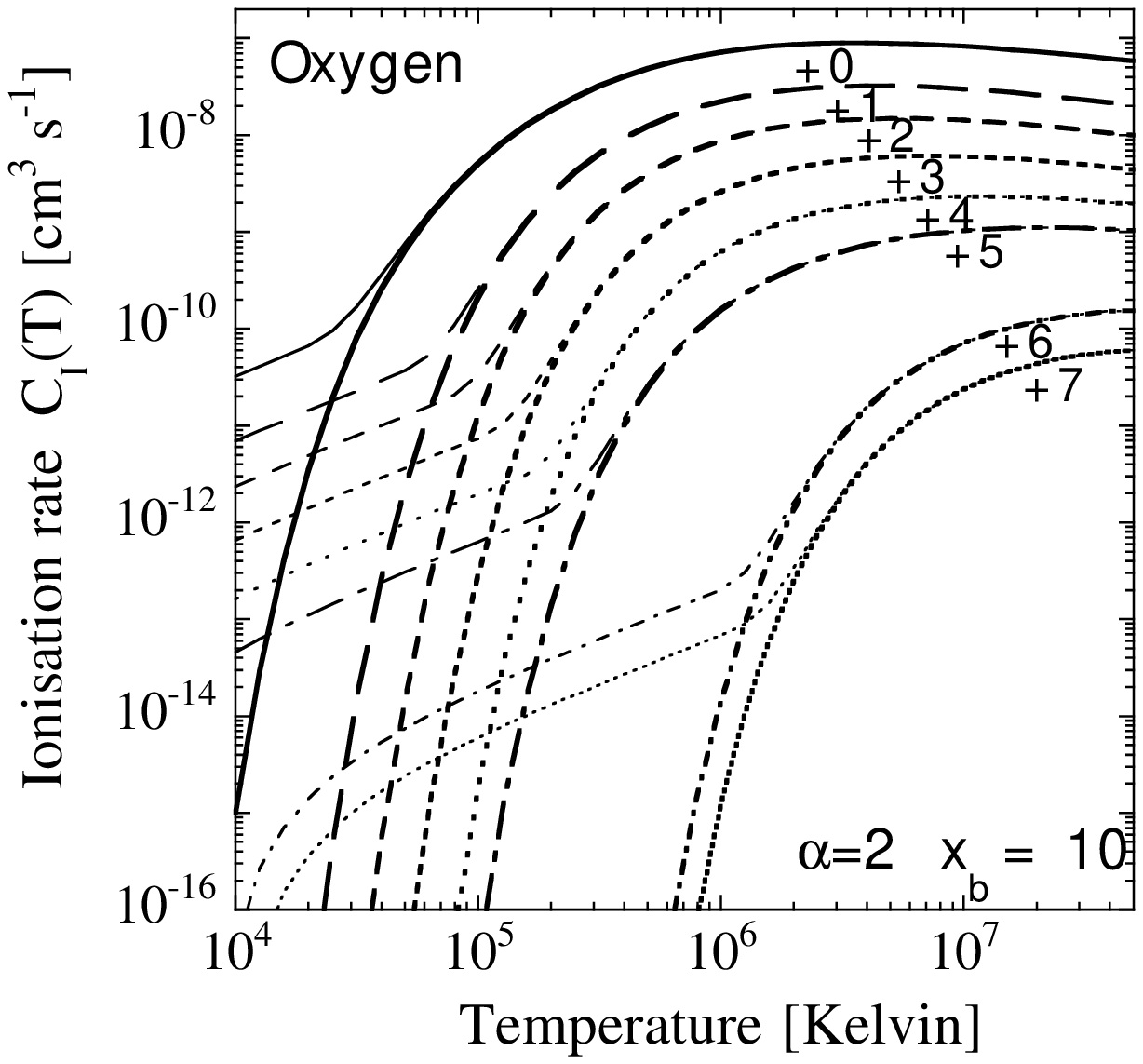}
\caption{Variation of the ionization rate with temperature.  Each
curve corresponds to an ion of oxygen and is labeled accordingly.
Black lines:
rates for a Hybrid electron distribution with $\a=2$ and $\xb=10$.
Black thick lines: rates for a Maxwellian distribution. }
\label{fig:fIonisOT10}
\end{figure}

\begin{figure}[t]
\epsfxsize=8.cm \epsfverbosetrue \epsfbox{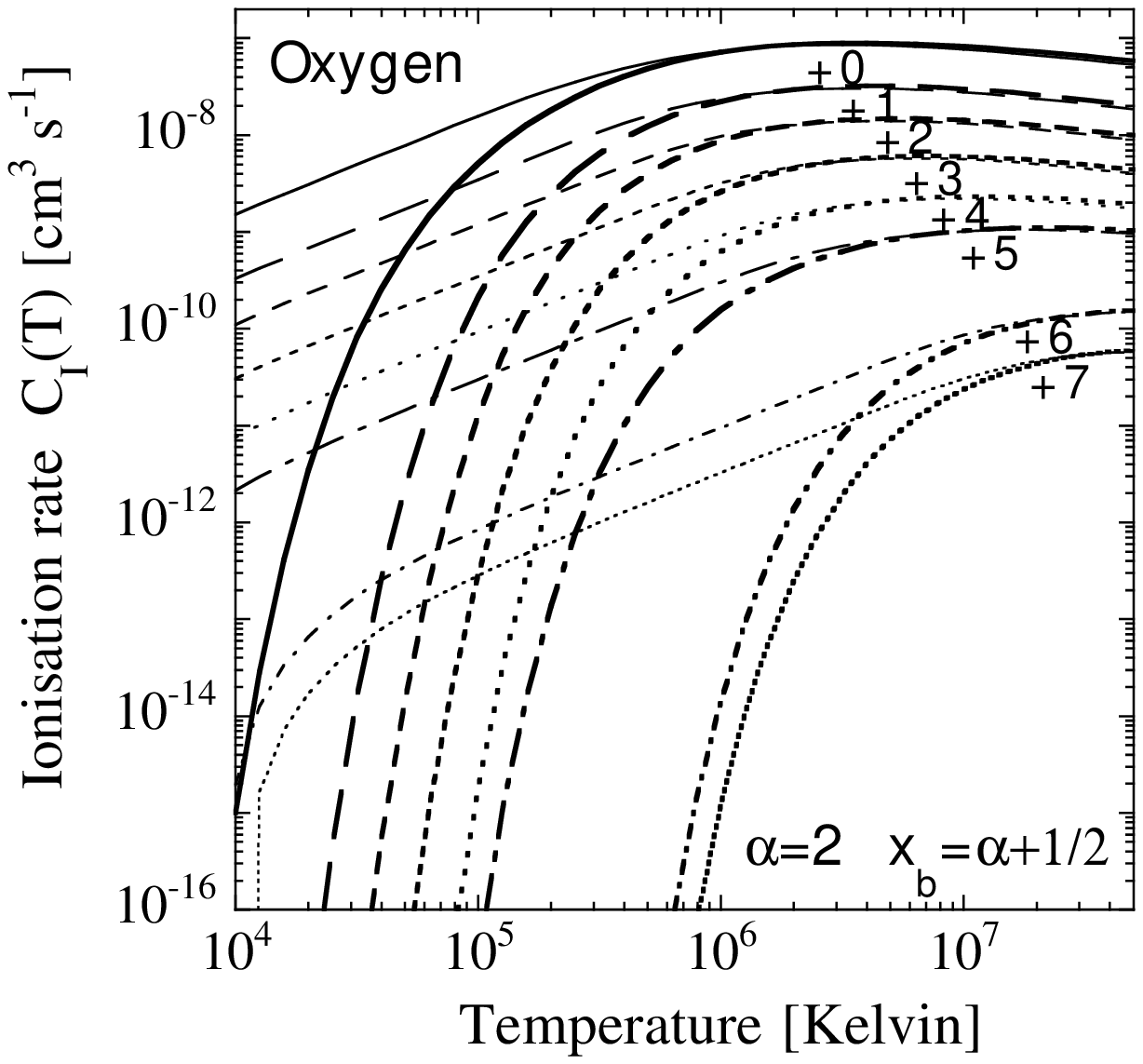}
\caption{Same as Fig.~\ref{fig:fIonisOT10} for $\a=2$ and
$\xb=\a+\frac{1}{2}=2.5$.}
\label{fig:fIonisOT2p5}
\end{figure}

The ionization rates for a Hybrid distribution are less dependent on
the temperature than the Maxwellian rates, as illustrated on
Fig.~\ref{fig:fIonisOT10} and on Fig.~\ref{fig:fIonisOT2p5}.  This is a
direct consequence of the temperature dependance of the enhancement
factor: as this factor increases with decreasing temperature, the
Hybrid ionization rate decreases less steeply with temperature than
the Maxwellian rates.  More precisely, as derived from the respective
expression of the rates at low temperature (respectively
Equation~\ref{eq:CDIM} and Equation~\ref{eq:GDIH2}),
the Maxwellian rate falls off exponentially (as $\e^{-\EI/kT}$) with
decreasing temperature, whereas the Hybrid rate only decreases as a
power--law.  As expected, one also notes that the modification of the
rates is more pronounced for lower value of $\xb$ (compare the two
figures corresponding to $\xb=10$ and $\xb=2.5$).

\subsection{The  recombination
rates}\label{sec:recombination}
\subsubsection{The radiative recombination rates}

The radiative recombination rates are expected to be less affected by
the Hybrid distribution, since the cross sections for recombination
decrease with energy and no threshold exists.  As the net effect
of the high energy tail present in the hybrid distribution is to
increase the median energy of the distribution (cf. fig.~\ref{fig:fCEmed}), as compared to a
Maxwellian, the radiative recombination rates are decreased.  \\

To estimate the corresponding dumping factor, $\bRR = \rrH/\rrM$, we 
follow the method used by Owocki \& Scudder (\cite{Owocki83}). We 
assume that the radiative recombination cross section varies as a 
power--law in energy:
\begin{equation}
\sRR\propto{E^{-a}}
\label{eq:sRR}
\end{equation}
which corresponds to a recombination rate (Equation~\ref{eq:C(kT)}), for a
Maxwellian
distribution (Equation~\ref{eq:distM}), varying as:
\begin{equation}
\rrM\propto{T^{\eta}}
\end{equation}
with $\eta = a - \frac{1}{2}$. \\

\begin{figure}[t]
\centering
\epsfxsize=8.cm \epsfverbosetrue \epsfbox{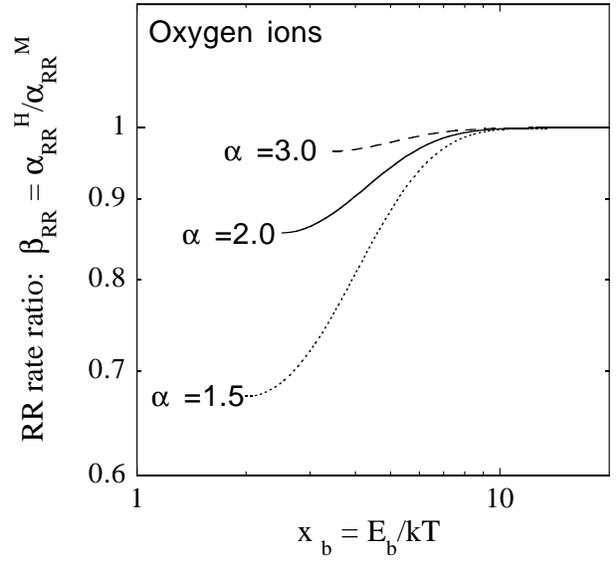} 
\caption{Dumping
factor for the radiative recombination rates, $\bRR$,
of oxygen ions versus the break parameter $\xb$.  The
curves correspond to different values of the slope parameter $\a$ and
are labeled accordingly. This dumping factor is independent of
temperature.}
\label{fig:fRROa}
\end{figure}

The dumping factor computed for such a power--law cross section
(Equation~\ref{eq:C(kT)} with Equation~\ref{eq:sRR}) is:
\begin{equation}
\bRR = \frac{ \int_{0}^{\infty} x^{-\eta}\ f^{\rm H}(x)\ dx}
{\int_{0}^{\infty} x^{-\eta}\ f^{\rm M}(x)\ dx}
\end{equation}
Note that the dumping factor is independent of the temperature. It 
depends on the ion considered via the $\eta$ parameter. Replacing the 
Maxwellian and Hybrid distribution functions by their expression 
(respectively Equations~\ref{eq:distM} and \ref{eq:distH}) we obtain:
\begin{equation}
\bRR = \frac{C(\xb,\a)}{\!\Gamma\!\left(\frac{3}{2}\!-\eta\right)}\!
\left[\gamma\!\left(\frac{3}{2}\!-\eta,\xb\right) +
\frac{\xb^{\frac{3}{2}\!-\eta}~e^{-\xb}}{\a+\!\eta\!-1}\right]
\label{eq:bRR}
\end{equation}

This estimate of the dumping factor is only an approximation, since
the radiative recombination has to be computed by summing over the
various possible states of the recombined ions, taking into account
the respective different cross sections.  Furthermore, even if often
the radiative recombination rate can be approximated by a power--law in
a given temperature range, this does not mean that the underlying
cross section is well approximated by a unique power--law.

However as we will see the correction factor is small, and we can
reasonably assume that it allows a fair estimate of the true Hybrid
radiative recombination rates.  To minimize the errors, the Hybrid
radiative recombination rate has to be calculated from the best
estimate of the Maxwellian rates, multiplied by this approximation of
the dumping factor:
\begin{equation}
 \rrH = \bRR\ \rrM
 \label{eq:aRRH}
\end{equation}
where $\rrM$ is as given in Mazzotta \etal (\cite{Mazzotta1998}).

The parameters $\eta$ for the various ions are taken from Aldrovandi
\& P{\'e}quignot (\cite{Aldrovandi73}), when available.  For other
ions we used a mean value of $\eta=0.8$ corresponding to the mean
value $<\eta>$ reported in Arnaud \& Rothenflug (\cite{Arnaud85}). 
The exact value has a negligible effect on the estimation of the
radiative recombination rates.\\

\begin{figure}[t]
\centering
\epsfxsize=8.cm \epsfverbosetrue \epsfbox{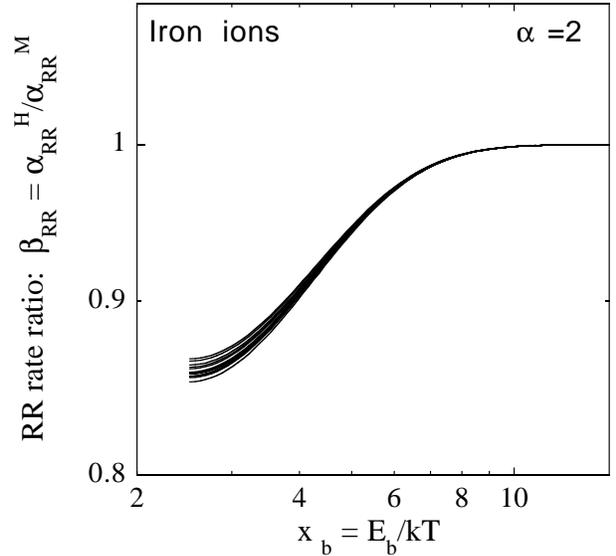}
\caption{Same as Fig.~\ref{fig:fRROa} for iron ions and $\a=2$.  The slight
variations among the species are due to the corresponding variations in the
adopted $\eta$ values (see Equation~\ref{eq:bRR}). }
\label{fig:fRRFea}
\end{figure}

The dumping factor is plotted in Fig.~\ref{fig:fRROa} for the various ions 
of oxygen.  In that case a common $\eta$ value is used.  The dumping 
factor decreases with decreasing values of $\xb$ and $\a$, following the 
increase of the distribution median energy. The modification is however 
always modest, at most $15\% $ for $\a=2$.  For iron, plotted in 
Fig.~\ref{fig:fRRFea} for $\a=2$, the value of $\eta$ slightly changes with 
the considered ions, but this only yields negligible variations in the 
dumping factor.

\subsubsection{The dielectronic recombination rates}

The dielectronic recombination is a resonant process involving bound
states at discrete energies $E_{i}$ and the rates have to be computed
by summing the contribution of many such bound states.  According to
Arnaud \& Raymond (\cite{Arnaud92}), and Mazzotta et al. 
(\cite{Mazzotta1998}), the dielectronic recombination rates for a
Maxwellian distribution can be fitted accurately by the formula:
\begin{equation}
\label{eq:aDRM}
\drM =T_{\rm eV}^{-3/2}\ \sum_{\rm i}c_{\rm i}~e^{-x_{\rm 
i}}~~~~~~~{\rm cm^{3}\ s^{-1}}
\end{equation}
where T$_{\rm eV}$ is the temperature expressed in eV and $x_{\rm i}=
E_{\rm i}/\kT$.  The numerical values for $c_{\rm i}$ and $E_{\rm i}$
are taken from Mazzotta et al.  (\cite{Mazzotta1998}).  Only a few
terms (typically 1 to 4) are introduced in this fitting formula.  They
roughly correspond to the dominant transitions for the temperature
range considered.  \\

\begin{figure}[t]
\centering
\epsfxsize=8.cm \epsfverbosetrue \epsfbox{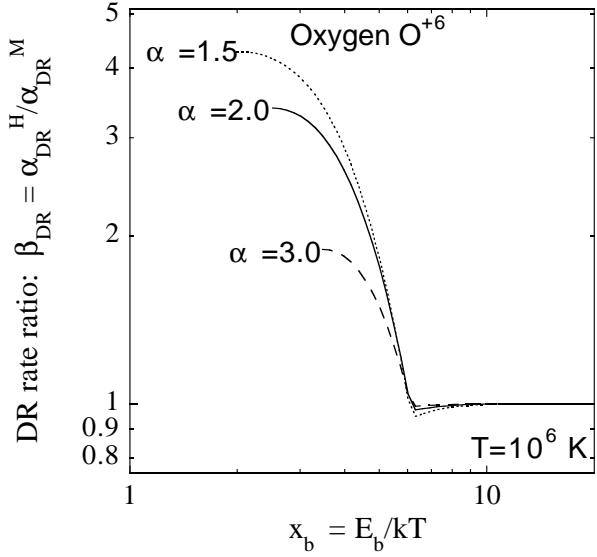}
\caption{Variation of the ratio of the DR rate in a Hybrid
distribution over that in a Maxwellian with the same temperature,
versus the break parameter $\xb$ of the Hybrid distribution.  The
curves correspond to different values of the slope parameter $\a$ and
are labeled accordingly.  The ion considered is O$^{+6}$, the
temperature is fixed at $10^{6}$~K. }
\label{fig:fDRO7a}
\end{figure}

Following again the method used by Owocki \& Scudder
(\cite{Owocki83}), we thus assume that the corresponding dielectronic
recombination cross section can be approximated by:
\begin{eqnarray}
\sDR&=&\sum_{\rm i} C_{\rm i}~\delta(E-E_{\rm i})~~{\rm with}~~ C_{\rm
i}=\frac{c_{\rm i}\ (2\pi \me)^{\frac{1}{2}}}{4~E_{\rm i}}
\label{eq:sDR}
\end{eqnarray}
The relation between $C_{\rm i}$ and $c_{\rm i}$ is obtained by
comparing Equation (\ref{eq:aDRM}) with the equation obtained by
integrating (Equation~\ref{eq:C(kT)}) the above cross section over a
Maxwellian distribution (Equation~\ref{eq:distM}).  The dielectronic 
rates can then be computed from Equation~(\ref{eq:C(kT)}), with the cross 
section given by Equation~(\ref{eq:sDR}) and the distribution function 
given by Equation~(\ref{eq:distH}):
\begin{eqnarray}
\label{eq:aDRH}
\drH& =&C(x_{\rm b},\a)\ T_{\rm eV}^{-3/2} \sum_{i,x_{\rm i} \leq x_{\rm b}}\!c_{\rm i}\ e^{-x_{\rm i}}  \\
&+& C(x_{\rm b},\a)\ T_{\rm eV}^{-3/2} e^{-x_{\rm b}}\!\sum_{i,x_{\rm i} >x_{\rm b}}\!c_{i}\left(\frac{x_{\rm i}}{x_{\rm b}}\right)^{-\a-\frac{1}{2}}\nonumber
\end{eqnarray}
Note that this estimate of $\drH$ is only an approximation, for the same
reasons outlined above for the radiative recombination rates.\\

To understand the effect of the hybrid distribution, let us assume
that only a single energy $\EDR$ is dominant, corresponding to a simple
Dirac cross section at this energy.  In that case, from 
Equation~(\ref{eq:C(kT)}), the ratio of the dielectronic recombination
rate in a Hybrid distribution over that in a Maxwellian with the same
temperature, $\bDR = \drH/\drM$, is simply the ratio of the Hybrid 
to the Maxwellian function at the resonance energy. Its expression depends 
on the position of the resonance energy with respect to the energy break.  
In reduced energy coordinates, we obtain from Equation~(\ref{eq:distM}) 
and Equation~(\ref{eq:distH}):
\begin{eqnarray}
\bDR & = &C(\xb,\a) ~~~~~~~~~\mathrm{for}~\frac{\EDR}{\kT} \leq \xb  \\
{\bDR} & = & C(\xb,\a)\
{\rm e}^{\left(\frac{\EDR}{\kT}-\xb\right)}\left(\frac{\EDR}{\xb \kT}\right)^{\!-(\frac{1}{2}
+\a)}\nonumber \\
&&~~~~~~~~~~~~~~~~~~~~~\mathrm{for}~\frac{\EDR}{\kT} \geq \xb \nonumber
\end{eqnarray}

\begin{figure}[t]
\centering
\epsfxsize=8.cm \epsfverbosetrue \epsfbox{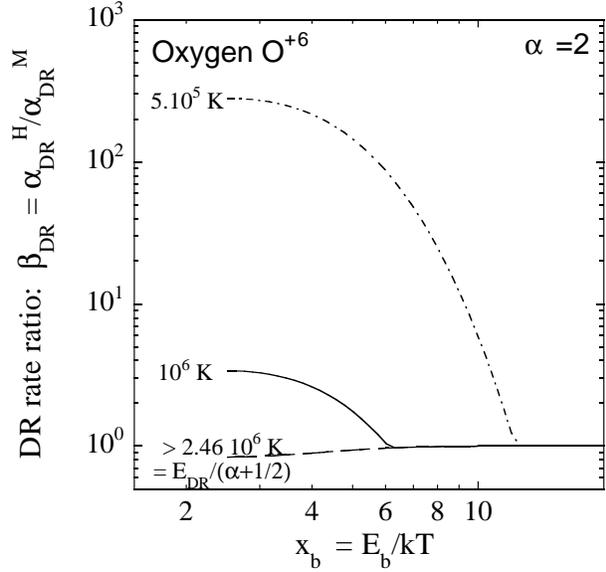}
\caption{Same as Fig.~\ref{fig:fDRO7a} but the parameter $\a$ is now
fixed to 2. The curves correspond to different values of the
temperature and are labelled accordingly.}
\label{fig:fDRO7T}
\end{figure}

\begin{figure}[t]
\centering
\epsfxsize=8.cm \epsfverbosetrue \epsfbox{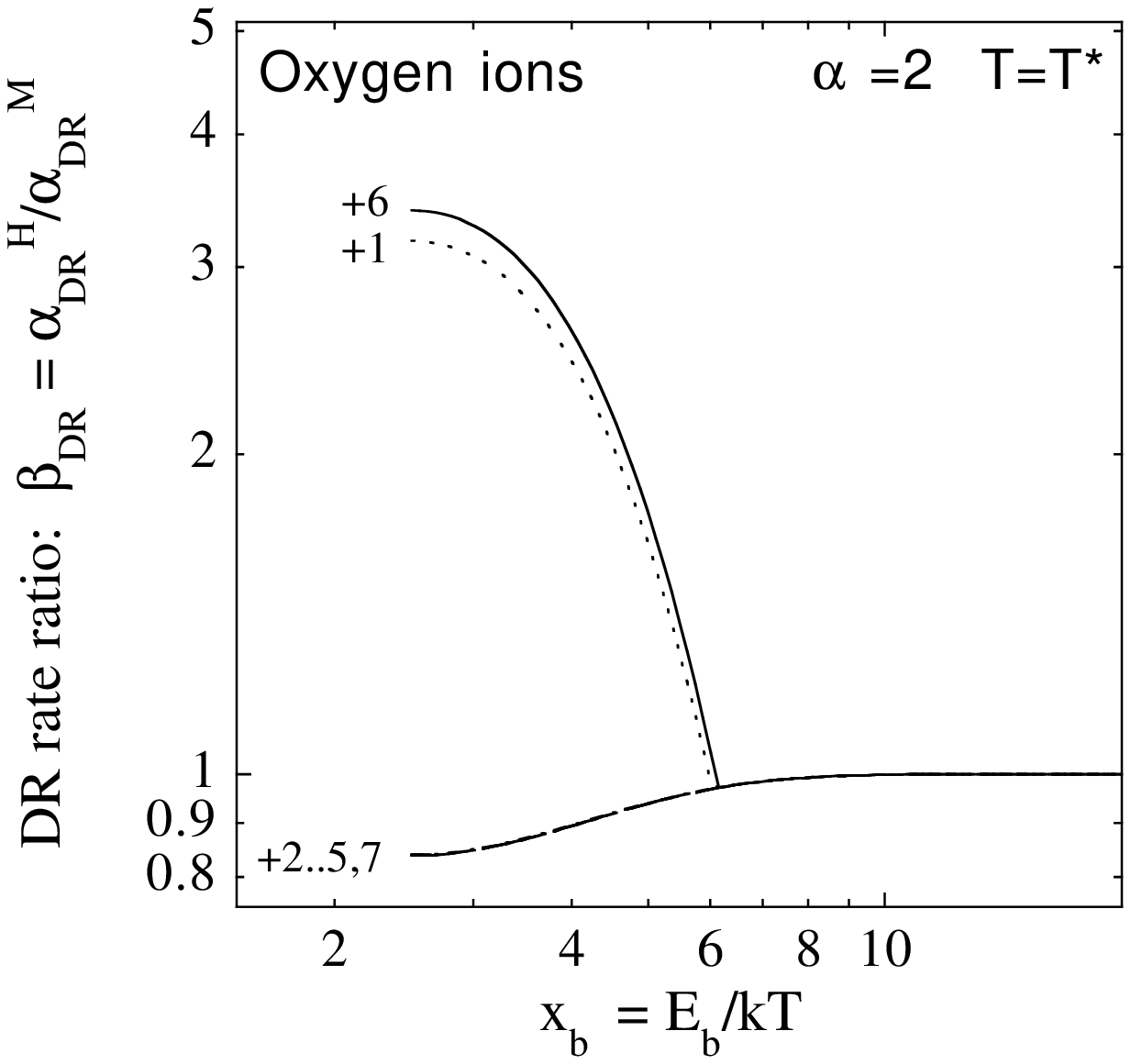} 
\caption{Same as Fig.~\ref{fig:fDRO7a} for the different ions of oxygen.  
Each curve is labeled by the charge of the ion considered.  The slope parameter is
fixed to $\a=2$ and the temperature for each ion is fixed at the
value, $T^{*}$, where the abundance of the ion is maximum for a
Maxwellian electron distribution.}
\label{fig:fDROTs}
\end{figure}

\begin{figure}[t]
\centering
\epsfxsize=8.cm \epsfverbosetrue \epsfbox{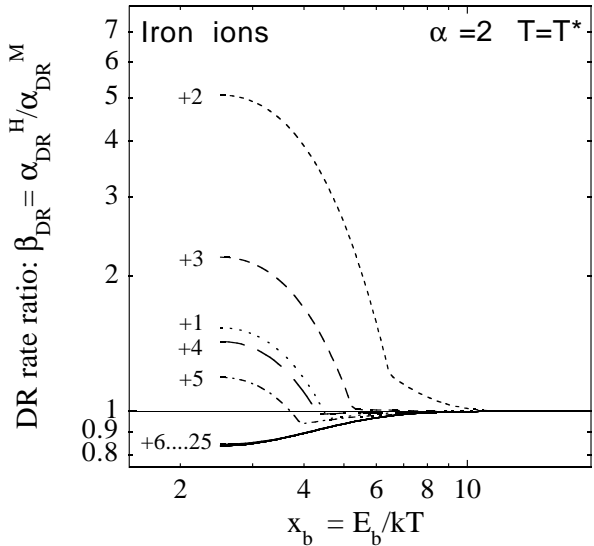}
\caption{Same as Fig.~\ref{fig:fDROTs} for iron.}
\label{fig:fDRFeTs}
\end{figure}

For $(\xb~\kT) > \EDR$ (i.e. at high temperature or high value of $\xb$), 
the resonance lies in the Maxwellian part of the distribution.  $\bDR$ 
is independent of the temperature and the dielectronic recombination 
rates are decreased, following the variation of the normalisation 
factor, $C(\xb,\a)$, i.e. the decrease is modest (see fig.~\ref{fig:fCEmed}).

For $(\xb~\kT) < \EDR$ the resonance lies in the power--law part of
the distribution.  The increase of the dielectronic recombination rate
can be dramatic, increasing with decreasing $\xb$ and $\a$.  \\

These effects of the Hybrid distribution on the dielectronic
recombination rates are illustrated in Fig.~\ref{fig:fDRO7a} to
Fig.~\ref{fig:fDRFeTs}, where we plotted the factor $\bDR$ for various
ions and values of the parameters.  The factors are computed exactly
from Equations~(\ref{eq:aDRM}) and (\ref{eq:aDRH}).

In Fig.~\ref{fig:fDRO7a} we consider O$^{+6}$ at the temperature of
its maximum ionization fraction, $T^*=10^{6}~{\rm K}$.  For this ion
only one term is included in the rate estimate, with $\EDR = 529~{\rm
eV}$, and $\EDR/\kT=6.1$ at the temperature considered.  We plotted
the variation of $\bDR$ with $\xb$ for $\a=3$, $\a=2$ and $\a=1.5$. 
For $\xb > 6.1$ the `resonance' energy $\EDR$ lies in the Maxwellian
part of the distribution and the dielectronic recombination rate is
decreased as compared to a Maxwellian, but by less than $10\%$, following the variation of the normalisation factor $C(\xb,\a)$.  For smaller values of $\xb$, the rate is increased significantly, up to a factor of 5 for $\a=1.5$.

We consider other temperatures, fixing $\a$ to $\a=2$, in
Fig.~\ref{fig:fDRO7T}.  Since we only consider the parameter range
$\xb > \a+1/2$, there is a threshold temperature, $kT >
\EDR/(\a+1/2)$, above which the resonance always falls in the
Maxwellian part.  The dielectronic recombination rate is decreased via
the factor $C(\xb,\a)$.  This factor slightly decreases with
decreasing $\xb$ (cf. fig.~\ref{fig:fCEmed}).  At lower temperature, the resonance energy can fall
above the break, provided that $\xb$ is small enough ($\xb<\EDR/\kT$).  
This occurs at smaller $\xb$ for higher temperature and the enhancement 
at a given $\xb$ increases with decreasing temperature.

We display in Fig.~\ref{fig:fDROTs} and Fig.~\ref{fig:fDRFeTs} the
variation of the factor $\bDR$ with $\xb$ (for $\a=2$), for the
different ions of oxygen and iron, at the temperature of maximum
ionization fraction for a Maxwellian distribution under ionization 
equilibrium.  For most of the ions this temperature is above the 
threshold temperature, $kT = \EDR/(\a+1/2)$, for all the resonances and 
the dielectronic rate is
decreased.  For the ions for which this is not the case (O$^{+1}$, 
O$^{+6}$ and from Fe$^{+1}$ to Fe$^{+5}$), the dielectronic rate can be 
increased significantly (by a factor between 2 to 5) provided $\xb$ is 
small enough (typically $\xb=2.5-5$).  The increase starts as soon as
$\xb<\EDR/\kT^*$ for the oxygen ions.  The behavior of $\bDR$ is more
complex for the iron ions (two breaks in the variation of $\bDR$), due
to the presence of more than one dominant resonance energy (more than one
term), taken into account in the computation of the dielectronic 
rate. \\
\begin{figure}
\epsfxsize=8.cm \epsfverbosetrue \epsfbox{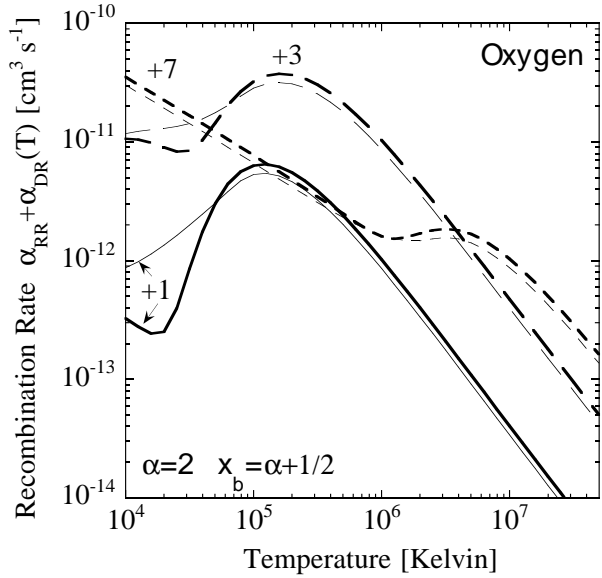}
\caption{Variation of the total recombination rate with temperature.  Each curve corresponds to an ion and is labeled accordingly.  Black lines: rates for a Hybrid electron distribution with $\a=2$ and  $\xb=\a+\frac{1}{2}=2.5$. Black thick lines: rates for Maxwellian distribution.}
\label{fig:fRtotOT2p5}
\end{figure}

In conclusion, the effect of the hybrid distribution on the
dielectronic rate depends on the position of the resonance energy as
compared to the power--law energy break.  It can only be
increased if $\kT < \EDR/(\a+1/2)$. At high temperature, the
dielectronic recombination rate is slightly decreased. 

\subsubsection{The total recombination rates}

At $\xb =10$, the
total rates are basically unchanged by the Hybrid distribution.  For
$\xb=2.5=\a+1/2$ (Fig.~\ref{fig:fRtotOT2p5}), the total rates are more
significantly changed.  The radiative recombination rate increases
with decreasing temperature and it usually dominates the total
recombination rate in the low temperature range.  As the dielectronic
rate is increased by the Hybrid distribution only at low temperature,
there are very few ions for which the total recombination rate can be
actually increased.  This only occurs in a small temperature range, in
the rising part of the dielectronic rate.  One also notes the expected
slight decrease of the radiative recombination rates (when it is
dominant at low temperature) and of the dielectronic rate at high
temperature.



\section{Ionization equilibria}\label{sec:Ionizationequilibria}

\begin{figure*}[t]
\epsfxsize=\textwidth \epsfverbosetrue \epsfbox{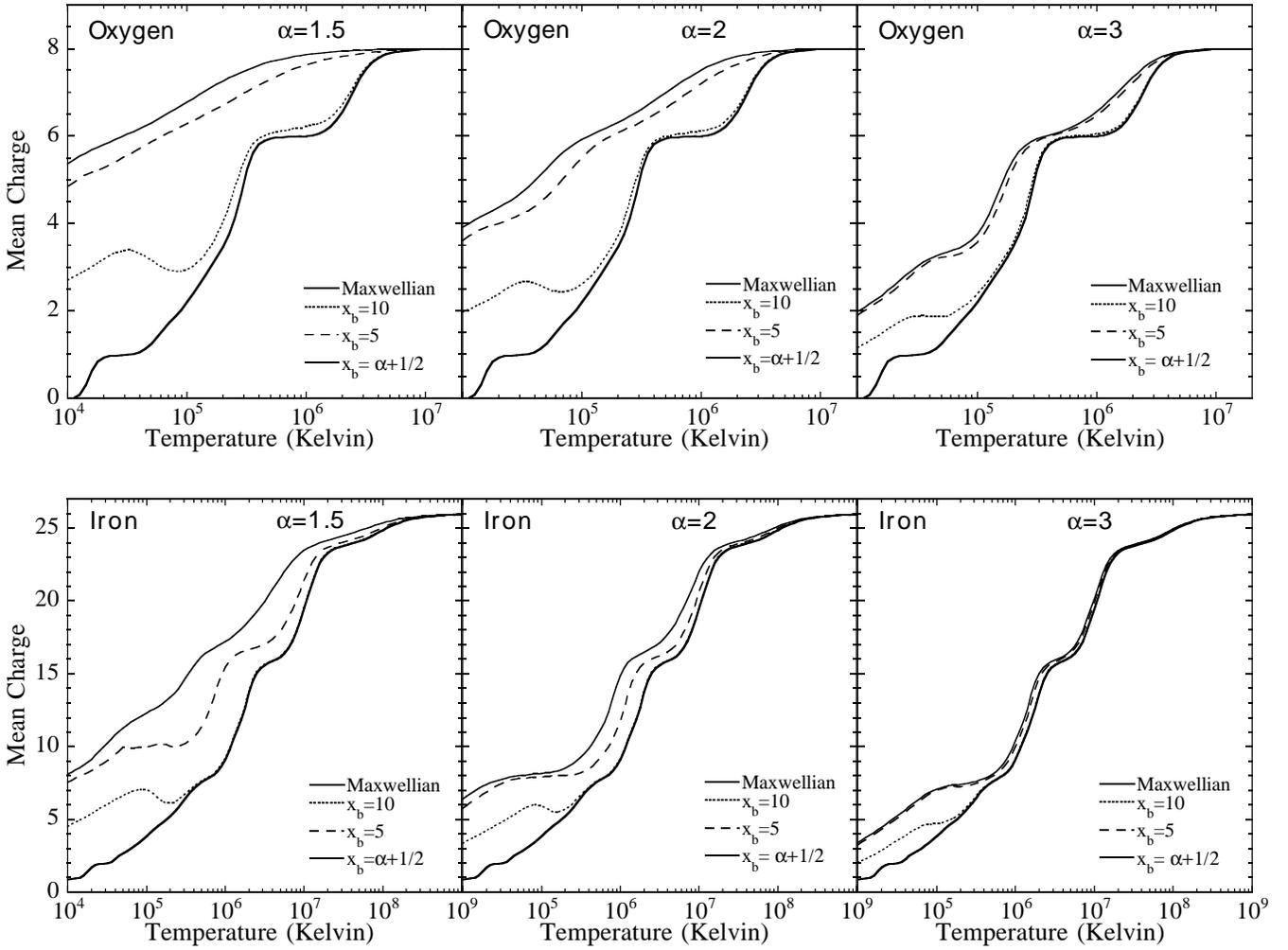}
\caption{Mean electric charge versus temperature for oxygen (top
panels) and iron (bottom panels) for different electron distributions. 
Black thick lines: Maxwellian electron distribution.  Black thin lines:
Hybrid distribution with $\a=1.5$ (left panels), $\a=2$ (middle
panels) and $\a=3$ (right panels).  Three values of $\xb$ are
considered: $\xb=10$ (dotted line), $\xb=5$ (dashed line) and the
extreme value $\xb=\a+1/2$ (full line).}
\label{fig:fZmeanOFe}
\end{figure*}

\subsection{Collisional ionization equilibrium (CIE)}

The ionization equilibrium fractions, for coronal plasmas, can be
computed from the rates described in the previous sections.  In the
low density regime (coronal plasmas) the steady state ionic fractions
do not depend on the electron density and the population density ratio
$N_{\rm Z,z+1}$/$N_{\rm Z,z}$ of two adjacent ionization stages
$Z^{+(z+1)}$ and $Z^{+z}$ of element $Z$ can be expressed by:
\begin{equation}
\frac{N_{\rm Z,z+1}}{N_{\rm Z,z}}=\frac{C^{\rm Z,z}_{\rm
I}}{\alpha^{\rm Z,z+1}_{\rm R}} 
\label{eq:frac}
\end{equation} where $C^{\rm Z,z}_{\rm I}$
and $\alpha^{\rm Z,z+1}_{\rm R}$ are the ionization and total
recombination rates of ion $Z^{+z}$ and $Z^{+(z+1)}$ respectively.
To assess the impact of the Hybrid rates on the ionization balance, we
computed the variation with temperature of the mean electric charge of
the plasma.  This variation is compared with the variation obtained
for a Maxwellian electron distribution on Fig.~\ref{fig:fZmeanOFe} for
oxygen and iron and for different values of the parameters $\xb$ and
$\a$.  \\

As expected, the plasma is always more ionized for a Hybrid electron
distribution than for a Maxwellian distribution.  The mean charge at a
given temperature is increased, since the enhancement of the
ionization rate is always much more important than a potential
increase of the dielectronic rate (e.g. compare
Fig.~\ref{fig:fIonisOTs} and Fig.~\ref{fig:fDROTs}).  The effect of
the Hybrid distribution on the plasma ionization state is thus
governed by the enhancement of the ionization rates.  The enhancement
of the plasma mean charge is more pronounced for smaller values of
$\xb$ and smaller values of $\a$ (Fig.~\ref{fig:fZmeanOFe}), following
the same behavior observed for the ionization rates (due to the
increasing influence of the high energy tail).  Similarly the effect
is more important at low temperature, and a clear signature of the
Hybrid distribution is the disappearance of the lowest ionization
stages, that cannot survive even at very low temperature.  For
instance, for $\a=2$ and the extreme corresponding value of
$\xb=\a+1/2$, the mean charge is already +4 for oxygen and +6 for iron
at $T = 10^{4}$~K. At high temperature, the mean charge can
typically be changed by a few units, the effect being more important
in the temperature range where the mean charge changes rapidly with
temperature in the Maxwellian case.
\begin{figure*}[t]
\epsfxsize=\textwidth \epsfverbosetrue \epsfbox{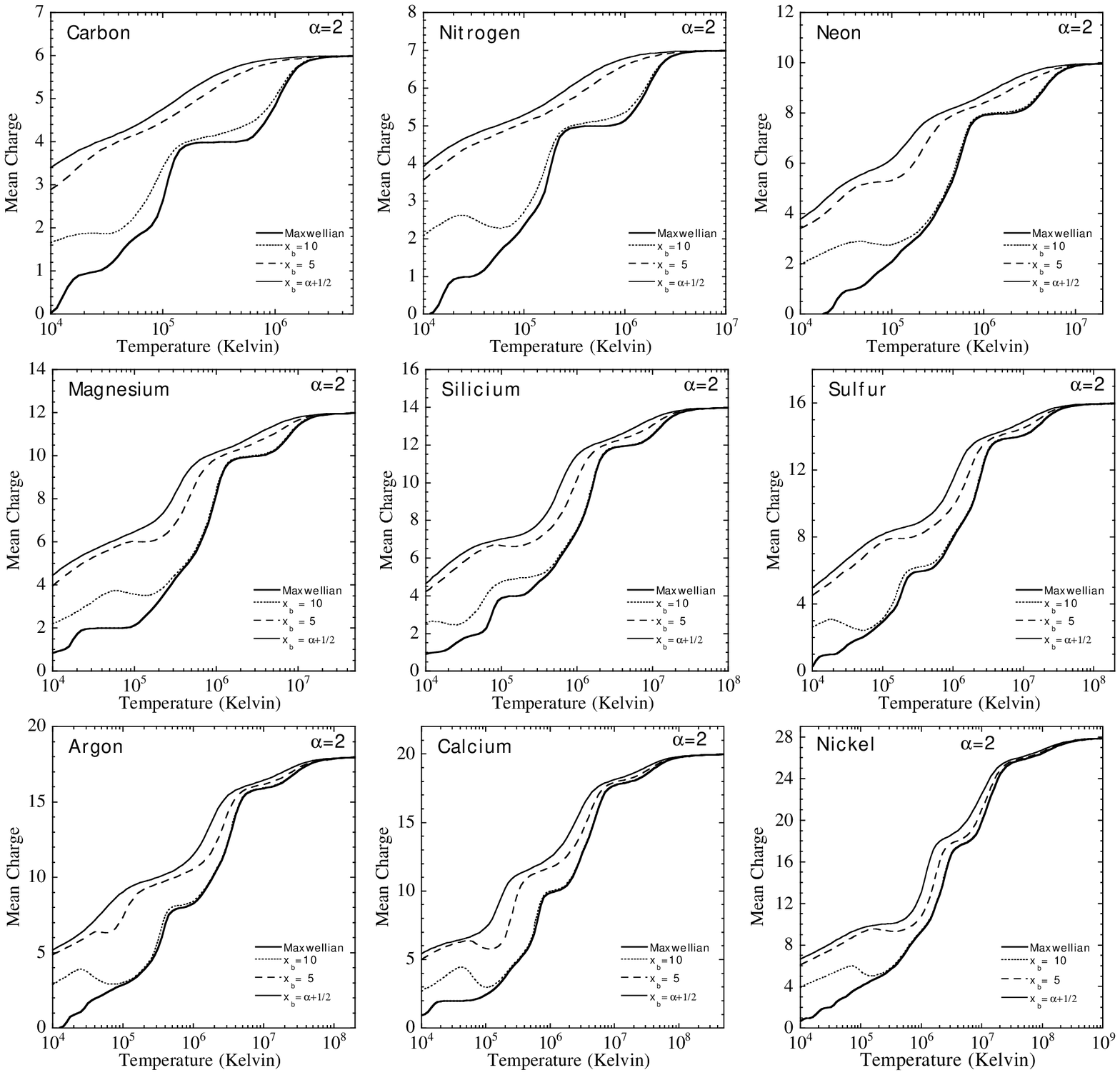}
\caption{Mean electric charge versus temperature for the Hybrid
electron distribution (black thin lines) compared to the Maxwellian
distribution (black thick lines).  The slope parameter $\a$ of the
Hybrid distribution is fixed to $2$, and the $\xb$ values are as in
Fig.~\ref{fig:fZmeanOFe}.  Each panel corresponds to an element and is
labelled accordingly.}
\label{fig:fZmean2}
\end{figure*}

The same behavior is seen for all elements (Fig.~\ref{fig:fZmean2}). 
One notes that the effect of the Hybrid distribution generally 
decreases with Z. Again this is a consequence of the same behavior 
observed on the ionization rates (see Fig.~\ref{fig:fIonisratio}). \\

A remarkable effect of the Hybrid distribution is that the mean charge
is not always a monotonous function of temperature, in the low 
temperature regime. This is clearly apparent in Fig.~\ref{fig:fZmeanOFe} 
and Fig.~\ref{fig:fZmean2} for $10^4~{\mathrm K} \le T \le 10^5$~K and 
$\xb=10$. This phenomenon can only occur when the
dielectronic rate dominates the total recombination rate and in the
temperature range where this rate increases with temperature.  In that
case, the density ratio of two adjacent ions, $N_{\rm Z,z+1}$/$N_{\rm
Z,z}$, can decrease with temperature provided that the ionization rate
of $Z^{+z}$ increases less rapidly with temperature than the
recombination rate of the adjacent ion $Z^{+(z+1)}$
(Eq.\ref{eq:frac}).  This usually does not occur in the Maxwellian
case, but can occur in the Hybrid case, due to the flatter temperature
dependence of the ionization rates for this type of distribution.  For
instance, for $ 3~10^{4}~\K \le T \le 7~10^{4}~\K$, the
ionization rate of O$^{+2}$ is increased by a factor of 2.5 for an
Hybrid distribution with $\xb=10$ (Fig.~\ref{fig:fIonisOT10}), whereas
the total recombination rate of O$^{+3}$ is increased by a slightly
larger factor of 2.7 (see the corresponding grey line in Fig.~\ref{fig:fRtotOT2p5}, as seen above for $x_{\rm b}$=10 the total rate is basically unchanged compared to the Maxwellian case).  The mean charge, which is
around $\langle z \rangle = 2.5$, is thus dominated by the behavior of 
these ions and decreases in that temperature range.

\subsection{Non--equilibrium ionization (NEI)}

\begin{figure*}[!ht]
\begin{center}
\resizebox{11.cm}{!}{\includegraphics{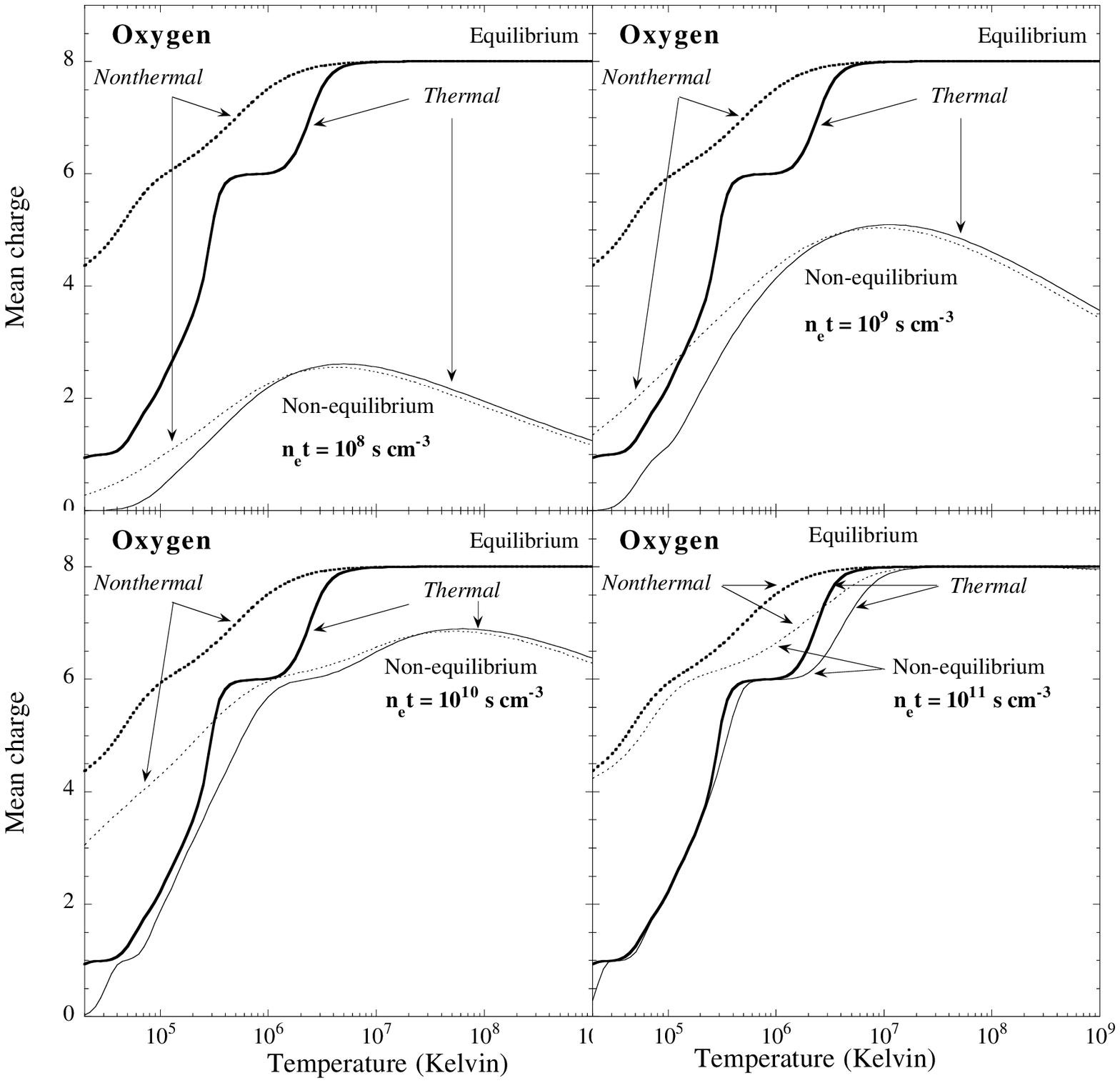}}\\
\resizebox{11.cm}{!}{\includegraphics{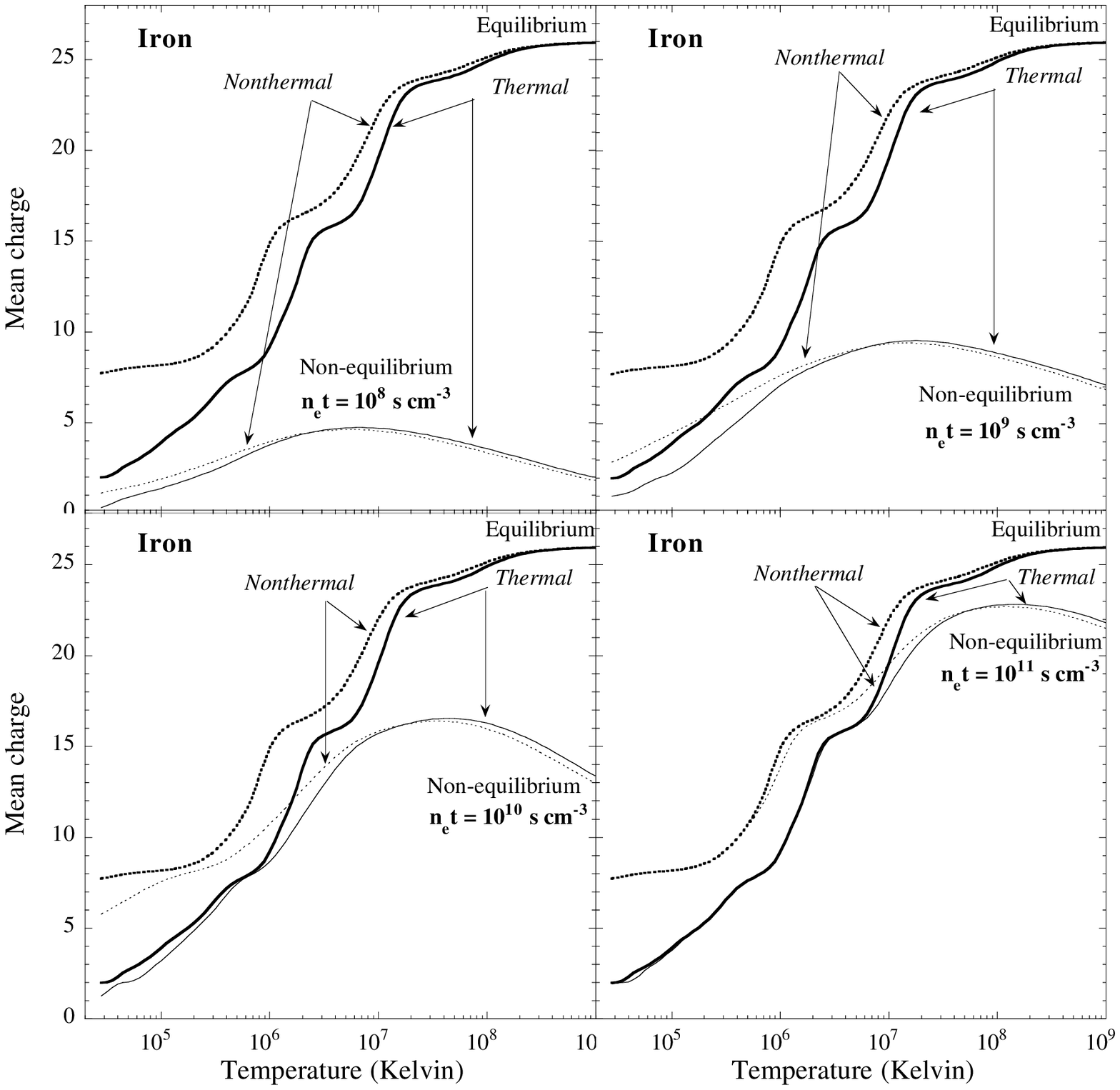}}
\caption{Mean electric charge versus temperature for oxygen (top panels) and 
iron (bottom panels) for different ionization timescales (thin lines) 
compared to equilibrium (bold lines) and for two extreme electron 
distributions, namely Maxwellian (solid lines) and Hybrid (dotted lines) for 
$\a = 2$ and $\xb = \a +1/2$.}
\label{fig:fZmeanNei}
\end{center}
\end{figure*}

Collisional Ionization Equilibrium (CIE) is not always achieved. For example, 
in adiabatic supernova remnants, the ionization timescale is longer than the 
dynamical timescale, so that the plasma is underionized compared to the 
equilibrium case. In non--equilibrium conditions, the ionization state of the 
gas depends on the thermodynamic history of the shocked gas (temperature, 
density) and time elapsed since it has been shocked. \\
The time evolution of the ionic fractions is given by:
\begin{eqnarray}
\frac{dX_{\rm Z,z}}{dt} &=& n_{\mathrm e}
[C_{\rm I}^{\rm Z,z-1}X_{\rm Z,z-1} 
+ {\alpha^{\rm Z,z+1}_{\rm R}}X_{\rm Z,z+1} \\ \nonumber
&-& (C_{\rm I}^{\rm Z,z} + {\alpha^{\rm Z,z}_{\rm R}})X_{\rm Z,z} ] \\
{\rm with}~&X_{\rm Z,z}& = \frac{N_{\rm Z,z}}{\!\sum_{\rm i}\! N_{\rm Z,i}}
\nonumber
\label{eq:fracNEI}
\end{eqnarray}
To estimate the effects of a Hybrid electron distribution on the ionization 
in non--equilibrium ionization conditions, we assume that the 
gas has been suddently heated to a given temperature, which stays constant 
during the evolution. The ionization timescale depends then on 
$\int{n_{\mathrm e}}~dt$, where $n_{\mathrm e}$ is the number density of 
electrons and $t$ the time elapsed since the gas has been heated. Within 
this assumption, the coupled system of rate equations can be resolved using 
an exponentiation method (e.g., Hughes \& Helfand 1985).

For different ionization timescales (up to equilibrium), we computed the 
variation with temperature of the mean electric charge of oxygen and iron in 
two extreme cases of the electron distribution: Maxwellian and Hybrid with 
$\xb=\a+1/2$ and $\a = 2$. For small ionization timescales ($n_{\mathrm e}~t 
\simeq 10^{8}-10^9$~s~cm$^{-3}$), the effect of the 
Hybrid distribution on the mean electric charge is small, it increases with 
the ionization timescale and is maximum at equilibrium as is illustrated for 
oxygen and iron in Fig.~\ref{fig:fZmeanNei}. As 
in the equilibrium case, the effect from non--thermal electrons is always more 
important at low temperature and vanishes at high temperature. Note that the 
mean electric charge is slightly larger at high temperature for the thermal 
population than for the non--thermal one, as a consequence of the decrease of 
the ionization cross section at very high energy.


\section{Conclusions}

We have studied the effect on the ionization and recombination 
rates, as well as on the ionization balance, of a non--thermal
electron distribution, as expected in the vicinity of strong shocks. 

The electron distribution is modelled by a Maxwellian distribution at
low energy up to a break energy, and by a power--law distribution at
higher energy.  It is caracterised by the three parameters $\kT$
(the temperature of the Maxwellian part), $\xb$ the reduced energy
break, and $\a$ the slope of the power--law component.  We only
considered the parameter range where $\xb > \a+1/2$ which corresponds
to an enhanced high energy tail.  All the behaviors outlined are only
valid for this range of parameters.

We provide exact formulae of the ionization rates for this Hybrid
electron distribution in the Appendix, and approximate estimates of the radiative recombination rates (Eq.~\ref{eq:bRR} and Eq.~\ref{eq:aRRH}) and of the dielectronic recombination rates (Eq.~\ref{eq:aDRH}).  The Hybrid rates depend on the ion considered and on the parameters $\kT$, $\a$ and $\xb$.  Computer codes are available on request.

For the parameter range considered, the proportion of electrons at high
energies and the mean energy of the distribution is a monotonic
function of $\xb$ and $\a$.  As expected, the modification of the
rates for the Hybrid distribution, as compared to the Maxwellian
distribution of the same temperature, increases with decreasing $\xb$
(with a threshold at about $\xb \sim 10-20$, higher for lower
temperature) and decreasing $\a$.

The impact of the Hybrid electron distribution on the ionization rates
depends on how the high energy tail affects the proportion of electrons
above the ionization potential $\EI$.  The Hybrid rates are increased
with respect to the Maxwellian rates except at very high temperature. 
The enhancement factor depends on the temperature, mostly via the
factor $\EI/\kT$, and increases dramatically with decreasing
temperature.  For a given ion, it is always important at $T^{*}$, the
temperature of maximum ionization fraction for a Maxwellian
distribution under ionization equilibrium, where it can reach several 
orders of magnitude.

The effect of the hybrid distribution on the dielectronic rate depends
on the position of the resonance energies $\EDR$ as compared to the
power--law energy break.  The dielectronic rate can only be increased
if $\kT < \EDR/(\a+1/2)$.  At $T^{*}$ the enhancement factor is
typically less than an order of magnitude.  At high temperature, the
dielectronic recombination rate is slightly decreased (by typically
$10\%$ at most). The effect of the hybrid distribution on the radiative 
recombination rates is only of the order of a few $10\% $ at most.

The ionization balance is affected significantly, whereas the effect
is smaller in ionizing NIE plasmas.  The plasma is always more ionized
for a Hybrid electron distribution than for a Maxwellian
distribution.  The effect is more important at low temperature, and a
clear signature of the Hybrid distribution is the disappearance of the
lowest ionization stages, which cannot survive even at very low
temperature.

\begin{acknowledgements}
We would like to thank Jean Ballet for a careful reading of the 
manuscript.
\end{acknowledgements}

\newpage
\appendix
\section{Ionization rates}\label{app:Ionization}

\subsection{Direct ionization (DI)}\label{app:DI}

For the direct ionization (DI) cross sections we chose the fitting
formula proposed by Arnaud \& Rothenflug (\cite{Arnaud85}) from the
work of Younger (\cite{Younger1981}):
\begin{eqnarray}\label{eq:QDI}
\sigma_{\DI}(E)\!&=&\!\sum_{j}\!\frac{1}{\uj \Ij^{2}}\!\left[\Aj~\Uj+ 
\Bj~\Uj^{2} +\Cj~\ln(\uj)+\Dj~\frac{\ln(\uj)}{\uj}\right]\nonumber \\
&&{\rm with}~~\uj=\frac{E}{\Ij}~; ~~~~\Uj=1-\frac{1}{\uj} 
\end{eqnarray}
The sum is performed over the subshells j of the ionizing ion.  E is
the incident electron energy and $\Ij$ is the collisional ionization
potential for the level $j$ considered.  \\
The parameters $\Aj$, $\Bj$, $\Cj$, $\Dj$ (in units of
10$^{-14}$\,cm$^{2}$\,eV$^{2}$) and $\Ij$ (in eV) are taken from the
works of Arnaud \& Raymond (\cite{Arnaud92}) for iron, and of Arnaud \&
Rothenflug (\cite{Arnaud85}) for the others elements.  The parameters
for elements not considered in these works are given in Mazzotta \etal
(\cite{Mazzotta1998}).

\subsubsection{The Maxwellian electron distribution}
For a Maxwellian electron distribution, Arnaud \& Rothenflug
(\cite{Arnaud85}) obtained according to Equations~(\ref{eq:fMaxw}),
(\ref{eq:C(kT)}) and (\ref{eq:QDI}), the rate:
\begin{equation}\label{eq:CDIM}
\CDIM =\frac{6.692\times 10^{7}}{(\kT)^{3/2}}
\sum_{j}\frac{{\rm e}^{-\xj}}{\xj} F^{\rm M}_{\DI}(\xj)~~~{\rm 
cm^3s^{-1}}
\label{eq:eq2a}
\end{equation}
where
\begin{eqnarray}
\label{eq:FDI}
\xj & =&\frac{\Ij}{\kT}\\
F^{\rm M}_{\DI}(\xj)&=& \Aj~[1-\xj f_{1}(\xj)]\nonumber \\
                    &+&\Bj~[1+\xj -\xj(2+\xj)f_1(\xj)]\\
                    &+&\Cj~f_1(\xj)+\Dj~\xj f_2(\xj)\nonumber
\end{eqnarray}

\noindent where $\kT$ and $\Ij$ are in eV. The summation is performed
over the subshells j of the ionizing ion.  The mathematical functions,
$f_1(x)={\rm e}^x \int_{1}^{\infty} \frac{{\rm e}^{-tx}}{t}~dt$, and $f_2(x)
={\rm e}^x \int_{1}^{\infty} \frac{{\rm e}^{-tx}}{t}~\ln (t)~dt$ can be computed
from the analytical approximations given by Arnaud \& Rothenflug
(\cite{Arnaud85}) in their Appendix\,B.

\subsubsection{The Hybrid electron distribution}

Similar to the Hybrid electron distribution, the direct ionization
rate $\CDIH$ is given by:
\begin{eqnarray}
\label{eq:CDIH}
\CDIH&=&C(\xb,\a)~\frac{6.692 \times 10^{7}}{(\kT)^{3/2}}\nonumber \\
&\times &\sum_{j}G^{\rm H}_{\DI}(\xj,\xb, \a)~~~{\rm 
cm^3s^{-1}} 
\end{eqnarray}
The function $G^{\rm H}_{\DI}(\xj,\xb, \a)$ depends on the position 
of the power--law break energy as compared to the ionization 
potential: 

\begin{itemize}
\item
For $\xb>\xj$:

$G^{\rm H}_{\DI}(\xj,\xb, \a)$ is the sum of the
contribution of the truncated Maxwellian component and the power--law
component:
\begin{eqnarray}
G^{\rm H}_{\DI}(\xj,\xb,\a)&=&\frac{{\rm e}^{-\xj}}{\xj}\ F^{\rm M}_{\DI}(\xj)-
\frac{{\rm e}^{-\xb}}{\xj}\ F^{'}_{\DI}(\xj,\xb) \nonumber\\ 
&+& {\rm e}^{-\xb}\ F_{\DI}^{\rm PL}(\ubj,\a)
\end{eqnarray}
where:
\begin{eqnarray}
    \ubj& =& \frac{\xb}{\xj} \\
F^{'}_{\DI}(\xj,\xb)&=&\Aj~\left[1-\xj f_{1}(\xb)\right]\nonumber \\
&+&\Bj~\left[1+\frac{\xj}{\ubj}-\xj (2+\xj) f_1(\xb)\right]\nonumber\\
&+&\Cj~\left[f_1(\xb) + \ln\left(\ubj\right)\right]\\
&+&\Dj~\left[\xj f_2(\xb)+\ln\left(\ubj\right) \xj f_1(\xb)\right] 
\nonumber\\
F_{\DI}^{\rm PL}(\ubj,\a)~&=& \Aj\left[ \frac{\ubj}{\a -1/2} -
\frac{1}{\a +1/2} \right]\nonumber \\
&+&\Bj\left[ \frac{\ubj}{\a -1/2} - \frac{2}{\a +1/2} +\frac{\ubj^{-1}}{\a +3/2} \right]\nonumber\\
&+&\frac {\Cj \ubj }{\a-1/2} \left[ \ln(\ubj) + \frac{1}{\a -1/2}\right] \\
&+& \frac{D_j}{\a+1/2} \left[\ln(\ubj) + \frac{1}{\a+1/2}\right] \nonumber
\end{eqnarray}

\item
For $\xb<\xj$:

Only the power--law component contributes of the electron distribution
to the rate:
\begin{eqnarray}
    \label{eq:GDIH2}
G^{\rm H}_{\DI}(\xj,\xb, \a)&=&{\rm e}^{-\xb}\ 
\left(\frac{\xb}{\xj}\right)^{\a+\frac{1}{2}}\ f_{\DI}(\a) 
\end{eqnarray}
where
\begin{eqnarray}
f_{\DI}(\a)&=&\frac{\Aj}{\a^{2}-1/4} + 
\frac{2 \Bj} {\left(\a^{2}-1/4\right)\left(\a+3/2\right)}\nonumber \\ 
&& + \frac{\Cj}{\left(\a-1/2\right)^{2}} 
+ \frac{\Dj}{\left(\a+1/2\right)^{2}}
\end{eqnarray}

\end{itemize}

\subsection{Excitation autoionization (EA)}\label{app:EA}

\noindent For the excitation autoionization (EA) cross sections, we
used the generalized formula proposed by Arnaud \& Raymond
(\cite{Arnaud92}):
\begin{eqnarray}\label{eq:QEA}
\sigma_{\EA}(E)\!&=&\!\frac{1}{u \IEA}\!\left[\!A+B~U+C~U_{2}+D~U_{3}+E~\ln(u)\right]\nonumber \\
&&{\rm with}~~u=\frac{E}{\IEA}~; ~~~~U_{n}=1-\frac{1}{u^n} 
\end{eqnarray}
where $\IEA$ is the excitation autoionization threshold and E is the 
incident electron energy.\\
The parameters $A$, $B$, $C$, $D$, $E$ (in units of
10$^{-16}$\,cm$^{2}$\,eV) and $\IEA$ (in eV) are taken from the works
of Arnaud \& Rothenflug (\cite{Arnaud85}) and Arnaud \& Raymond
(\cite{Arnaud92}).  The parameters for elements not considered in
these works are given in Mazzotta \etal (\cite{Mazzotta1998}).

\subsubsection{The Maxwellian electron distribution}
The excitation autoionization rate for a Maxwellian distribution is:
\begin{eqnarray}
\label{eq:CEAMaxw}
\CEAM&=&\frac{6.692 \times
10^{7}~{\rm e}^{-\xEA}}{(\kT)^{1/2}}~ F^{\rm M}_{\EA}(\xEA)~~~{\rm {cm}^3\,s^{-1}}
\end{eqnarray}
where
\begin{eqnarray}
 \label{eq:FEA}
\xEA&=&\frac{\IEA}{\kT}\\
F^{\rm M}_{\EA}(\xEA)&=& A+B[1-\xEA f_1(\xEA)]\nonumber \\
&+&C[1-\xEA+\xEA^{2} f_1(\xEA)]\\
&+&D\left[1-\frac{\xEA}{2}+\frac{\xEA^{2}}{2}-\frac{\xEA^3}{2} f_1(\xEA)\right]\nonumber\\
&+& E f_1(\xEA)\nonumber
\end{eqnarray}

\subsubsection{The Hybrid electron distribution}

For the Hybrid electron distribution, the excitation autoionization
rate $\CEAH$ is given by:
\begin{eqnarray}
\label{eq:CEA_H}
\CEAH&=&C(\xb,\a)~\frac{6.692 \times
10^{7}}{(\kT)^{1/2}}\nonumber \\
& \times &G^{\rm H}_{\EA}(\xEA,\xb, \a)~~~~~~~~{\rm cm^3s^{-1}} 
\end{eqnarray}
The function $G^{\rm H}_{\EA}(\xEA,\xb, \a)$ depends on the
position of the power--law break energy as compared to the ionization
potential: \\

\begin{itemize}
\item
For $\xb>\xEA$:

$G^{\rm H}_{\EA}(\xEA,\xb, \a)$ is the sum of the
contribution of the truncated Maxwellian component and the power--law
component:
\begin{eqnarray}
G^{\rm H}_{\EA}(\xEA,\xb,\a)&=&{\rm e}^{-\xEA}~F^{\rm M}_{\EA}(\xEA)\nonumber \\
&-& {\rm e}^{-\xb}~F^{'}_{\EA}(\xEA,\xb)\\
&+&\xb~{\rm e}^{-\xb}~F^{\rm PL}_{\EA}(\ucEA,\a)\nonumber
\end{eqnarray}
where:
\begin{eqnarray}
    \ucEA& =& \frac{\xb}{\xEA} \\
F^{'}_{\EA}(\xEA,\xb)&=& A+B\left[1-\xEA~f_1(\xb)\right]\nonumber \\
&+&C\left[1-\frac{\xEA}{\ucEA}+\xEA^{2}~f_1(\xb)\right] \\
&+&D\left[1-\frac{\xEA}{2 \ucEA^{2}}+
\frac{\xEA^{2}}{2 \ucEA}-\frac{\xEA^3}{2}~f_1(\xb)\right]\nonumber\\
&+&E\left[ \ln(\ucEA)+f_1(\xb)\right]\nonumber \\
F_{\EA}^{\rm PL}(\ucEA,\a)&=&A\left[\frac{1}{\a-1/2}\right]\nonumber \\
&+&B\left[\frac{1}{\a  -1/2} - \frac{\ucEA^{-1}}{\a +1/2}\right]\nonumber\\
&+&C\left[\frac{1}{\a  -1/2} - \frac{\ucEA^{-2}}{\a +3/2}\right]\\
&+&D\left[\frac{1}{\a  -1/2} - \frac{\ucEA^{-3}}{\a 
+5/2}\right]\nonumber \\
&+&E\left[\frac{1}{(\a -1/2)^2} +\frac{\ln(\ucEA)}{\a -1/2}\right]\nonumber
\end{eqnarray}

\item
For $\xb< \xEA$:

Only the power--law component contributes of the electron distribution
to the rate:
\begin{eqnarray}
G^{\rm H}_{\EA}(\xEA,\xb,\a)\!&=&\!\xb\
{\rm e}^{-\xb}\left(\frac{\xb}{\xEA} \right)^{\a-\frac{1}{2}}\!f_{\EA}(\a)
\end{eqnarray}
where
\begin{eqnarray}
f_{\EA}(\a)&=&\frac{A}{\a -1/2}+\frac{B}{\a^{2} 
-1/4}\nonumber\\
&+&\frac{2 ~C}{(\a -1/2)(\a  +3/2)}\\
&+&\frac{3~ D}{(\a  -1/2)(\a+5/2)}\nonumber\\
&+&\frac{E}{(\a-1/2)^2}\nonumber
\end{eqnarray}

\end{itemize}

\subsection{Total ionization rates (DI + EA)}

The total ionization  rate $\CIH$
 is obtained by:
\begin{eqnarray}
\CIH&=&\CDIH+\CEAH
\end{eqnarray}\end{document}